\documentclass{nature}

\usepackage{graphicx}
\usepackage{multirow}
\graphicspath{ {images/} }
\DeclareGraphicsExtensions{.png,.pdf,.eps}

\usepackage{hhline}
\usepackage{epstopdf}
\usepackage[square,sort,comma,numbers]{natbib}
\usepackage{amsmath,amssymb}

\usepackage[colorlinks,urlcolor=cyan,citecolor=blue,linkcolor=blue]{hyperref}

\newcommand{\aj}{Astron. J.}

\newcommand{\apj}{Astrophys. J.}
\newcommand{\apjl}{Astrophys. J., Letters}

\newcommand{\aap}{Astron. Astrophys.}

\newcommand{\mnras}{Mon. Not. R. Astron. Soc.}

\newcommand{\pasp}{Publ. Astron. Soc. Pacific}

\newcommand{\psj}{Planetary Science Journal}

\newcommand{\ssr}{Space Science Reviews}

\newcommand{\nat}{Nature}

\usepackage{xcolor}

\begin{document}

\title{A roadmap for the atmospheric characterization of terrestrial exoplanets with JWST}

\author{
\textbf{TRAPPIST-1 JWST Community Initiative},
Julien de Wit$^{\ref{miteaps},*}$,
Ren\'e Doyon$^{\ref{udem},\ref{trot},*}$, 
Benjamin V.\ Rackham$^{\ref{miteaps},\ref{mitkav}}$, 
Olivia Lim$^{\ref{udem},\ref{trot}}$, 
Elsa Ducrot$^{\ref{paris_region},\ref{cea}}$, 
Laura Kreidberg$^{\ref{mpia}}$, 
Bj\"{o}rn Benneke$^{\ref{udem},\ref{trot}}$,
Ignasi Ribas$^{\ref{ICECSIC},\ref{IEEC}}$, 
David Berardo$^{\ref{miteaps}}$,
Prajwal Niraula$^{\ref{miteaps}}$,
Aishwarya Iyer$^{\ref{ASU}}$,
Alexander Shapiro$^{\ref{mpis}}$,
Nadiia Kostogryz$^{\ref{mpis}}$,
Veronika Witzke$^{\ref{mpis}}$,
Micha\"{e}l Gillon$^{\ref{uliege1}}$, 
Eric Agol$^{\ref{washAA},\ref{washN}}$, 
Victoria Meadows$^{\ref{washAA},\ref{washN}}$, 
Adam J. Burgasser$^{\ref{san_diego}}$, 
James E. Owen$^{\ref{ICL}}$,
Jonathan J. Fortney$^{\ref{UCSC}}$,
Franck Selsis$^{\ref{UBord}}$,
Aaron Bello-Arufe$^{\ref{JPL}}$,
Zo\"{e} de Beurs$^{\ref{miteaps}}$,
Emeline Bolmont$^{\ref{ObsGen},\ref{CVGen}}$,
Nicolas Cowan$^{\ref{mcgillE},\ref{mcgillP}}$, 
Chuanfei Dong$^{\ref{BU}}$,
Jeremy J. Drake$^{\ref{CfA}}$,
Lionel Garcia$^{\ref{uliege1}}$,
Thomas Greene$^{\ref{NASAAmes}}$,
Thomas Haworth$^{\ref{QMUL}}$,
Renyu Hu$^{\ref{JPL},\ref{CalTechG}}$,
Stephen R. Kane$^{\ref{UCR}}$,
Pierre Kervella$^{\ref{LESIA}}$,
Daniel Koll$^{\ref{PekU}}$,
Joshua Krissansen-Totton$^{\ref{UWE}}$,
Pierre-Olivier Lagage$^{\ref{cea}}$,
Tim Lichtenberg$^{\ref{UGron}}$,
Jacob Lustig-Yaeger$^{\ref{JHU}}$,
Manasvi Lingam$^{\ref{FIT},\ref{UTA}}$,
Martin Turbet$^{\ref{UBord},\ref{Sorbo}}$, 
Sara Seager$^{\ref{miteaps},\ref{mitkav},\ref{mitAA}}$,
Khalid Barkaoui$^{\ref{uliege1},\ref{miteaps},\ref{IAC}}$,
Taylor J. Bell$^{\ref{NASAAmesB},\ref{NASAAmesSSAD}}$,
Artem Burdanov$^{\ref{miteaps}}$,
Charles Cadieux$^{\ref{trot}}$,
Benjamin Charnay$^{\ref{LESIA}}$,
Ryan Cloutier$^{\ref{McMU}}$,
Neil J. Cook$^{\ref{trot}}$,
Alexandre C. M. Correia$^{\ref{CFisUC},\ref{IMCCE}}$,
Lisa Dang$^{\ref{trot}}$,
Tansu Daylan$^{\ref{Princ}}$,
Laetitia Delrez$^{\ref{uliege1},\ref{uliege2}}$,
Billy Edwards$^{\ref{SRON}}$,
Thomas J. Fauchez$^{\ref{NASAGo},\ref{AmUn}}$,
Laura Flagg$^{\ref{Corn}}$,
Federico Fraschetti$^{\ref{CfA},\ref{LPL}}$,
Jacob Haqq-Misra$^{\ref{BlueM}}$,
Ziyu Huang$^{\ref{BU}}$,
Nicolas Iro$^{\ref{DLR}}$,
Ray Jayawardhana$^{\ref{Corn}}$,
Emmanuel Jehin$^{\ref{uliege2}}$,
Meng Jin$^{\ref{Lockh}}$,
Edwin Kite$^{\ref{UChi}}$,
Daniel Kitzmann$^{\ref{UBern}}$,
Quentin Kral$^{\ref{LESIA}}$,
David Lafrenière$^{\ref{trot}}$,
Anne-Sophie Libert$^{\ref{UNam}}$,
Beibei Liu$^{\ref{UZhej}}$,
Subhanjoy Mohanty$^{\ref{ICL}}$,
Brett M. Morris$^{\ref{STScI}}$,
Catriona A. Murray$^{\ref{UCol}}$,
Caroline Piaulet$^{\ref{trot}}$,
Francisco J. Pozuelos$^{\ref{IAA}}$,
Michael Radica$^{\ref{trot}}$,
Sukrit Ranjan$^{\ref{LPL}}$,
Alexander Rathcke$^{\ref{DTU}}$,
Pierre-Alexis Roy$^{\ref{trot}}$,
Edward W. Schwieterman$^{\ref{UCR}}$,
Jake D. Turner$^{\ref{Corn}}$,
Amaury Triaud$^{\ref{UBirm}}$,
Michael J. Way$^{\ref{NASAGISS},\ref{UUpp}}$
}

\maketitle

\textsl{$^*$These authors contributed equally to this work.}

\begin{affiliations}
\begin{footnotesize}
    \item Department of Earth, Atmospheric and Planetary Science, Massachusetts Institute of Technology, 77 Massachusetts Avenue, Cambridge, MA 02139, USA \label{miteaps}
    \item {Department of Physics, Universit\'{e} de Montr\'{e}al, Montreal, QC, Canada} \label{udem}
    \item {Trottier Institute for Research on Exoplanets, Universit\'{e} de Montr\'{e}al, Montreal, QC, Canada} \label{trot}
    \item Kavli Institute for Astrophysics and Space Research, Massachusetts Institute of Technology, Cambridge, MA 02139, USA \label{mitkav}
    \item Paris Region Fellow, Marie Sklodowska-Curie Action\label{paris_region}
    \item AIM, CEA, CNRS, Universit\'e Paris-Saclay, Universit\'e de Paris, F-91191 Gif-sur-Yvette, France \label{cea}
    \item {Max-Planck-Institut f\"ur Astronomie, K\"onigstuhl 17, D-69117 Heidelberg, Germany} \label{mpia}
    \item Institut de Ciències de l’Espai (ICE, CSIC), Campus UAB, c/ Can Magrans s/n, 08193 Bellaterra, Barcelona, Spain \label{ICECSIC}
    \item Institut d’Estudis Espacials de Catalunya (IEEC), c/ Gran Capità 2–4, 08034 Barcelona, Spain \label{IEEC}
    \item {School of Earth and Space Exploration, Arizona State University, 525 E. University Dr., Tempe AZ 85281} \label{ASU}
    \item Max Planck Institute for Solar System Research, Göttingen, Germany \label{mpis}

    \item Astrobiology Research Unit, Universit\'e de Li\`ege, 19C All\'ee du 6 Ao\^ut, 4000 Li\`ege, Belgium \label{uliege1}
    \item{Department of Astronomy and Astrobiology Program, University of Washington, Box 351580, Seattle, Washington 98195, USA} \label{washAA}
    \item{NASA Nexus for Exoplanet System Science, Virtual Planetary Laboratory Team, Box 351580, University of Washington, Seattle, Washington 98195, USA} \label{washN}
    \item Center for Astrophysics and Space Sciences, UC San Diego, UCSD Mail Code 0424, 9500 Gilman Drive, La Jolla, CA 92093-0424, USA \label{san_diego}
    \item Astrophysics Group, Imperial College London, Blackett Laboratory, Prince Consort Road, London SW7 2AZ, UK \label{ICL}
    \item Department of Astronomy and Astrophysics, University of California, Santa Cruz, Santa Cruz, 95064, CA, USA. \label{UCSC}
    \item {Laboratoire d'astrophysique de Bordeaux, Univ. Bordeaux, CNRS, B18N, all{\'e}e Geoffroy Saint-Hilaire, 33615 Pessac, France} \label{UBord}

    \item Jet Propulsion Laboratory, California Institute of Technology, Pasadena, CA, USA \label{JPL}
    \item Observatoire astronomique de l'université de Genève, Chemin Pegasi 51, CH-1290, Versoix, Switzerland\label{ObsGen}
    \item Centre Vie dans l’Univers, Faculté des sciences, Université de Genève, Quai Ernest-Ansermet 30, 1211 Genève 4, Switzerland \label{CVGen}
    \item Department of Earth \& Planetary Sciences, McGill University, 3450 rue University, Montréal, QC H3A 0E8, Canada \label{mcgillE}
    \item Department of Physics, McGill University, 3600 rue University, Montréal, QC H3A 2T8, Canada \label{mcgillP}
    \item Department of Astronomy, Boston University, Boston, Massachusetts 02215, USA\label{BU}
    \item Center for Astrophysics | Harvard \& Smithsonian, 60 Garden Street, Cambridge MA 02138\label{CfA}
    \item NASA Ames Research Center, Space Science and Astrobiology Division, MS 245-6, Moffett Field, CA, 94035 USA\label{NASAAmes}
    \item Astronomy Unit, School of Physics and Astronomy, Queen Mary University of London, London E1 4NS, UK \label{QMUL}
    \item Division of Geological and Planetary Sciences, California Institute of Technology, Pasadena, CA, USA. \label{CalTechG}
    \item Department of Earth and Planetary Sciences, University of California, Riverside, CA, USA \label{UCR}
    \item LESIA, Observatoire de Paris, Université PSL, CNRS, Sorbonne Université, Université de Paris, 5 place Jules Janssen, 92195 Meudon, France \label{LESIA}
    \item Department of Atmospheric and Oceanic Sciences, Peking University, Beijing, China \label{PekU}
    \item Department of Earth and Space Sciences/Astrobiology Program, University of Washington, Seattle, WA 98195, USA. \label{UWE}
    \item Kapteyn Astronomical Institute, University of Groningen, P.O. Box 800, NL-9700 AV Groningen, The Netherlands \label{UGron}
    \item The Johns Hopkins University Applied Physics Laboratory, 11100 Johns Hopkins Rd, Laurel, MD, 20723, USA \label{JHU}
    \item Department of Physics and Institute for Fusion Studies, The University of Texas at Austin, Austin, TX 78712, USA\label{UTA}
    \item Laboratoire de Météorologie Dynamique/IPSL, CNRS, Sorbonne Université, Ecole Normale Supérieure, Université PSL, Ecole Polytechnique, Institut Polytechnique de Paris, 75005 Paris, France ; \label{Sorbo}
    \item Department of Aeronautics and Astronautics, MIT, 77 Massachusetts Avenue, Cambridge, MA 02139, USA \label{mitAA}

    \item Instituto de Astrofísica de Canarias (IAC), Calle Vía Láctea s/n, 38200, La Laguna, Tenerife, Spain \label{IAC}
    \item BAER Institute, NASA Ames Research Center, Moffet Field, CA, USA \label{NASAAmesB}
    \item Space Science and Astrobiology Division, NASA Ames Research Center, Moffett Field, CA, USA\label{NASAAmesSSAD}
    \item Department of Physics \& Astronomy, McMaster University, 1280 Main St W, Hamilton, ON, L8S 4L8, Canada\label{McMU}
    \item CFisUC, Departamento de F\'isica, Universidade de Coimbra, 3004-516 Coimbra, Portugal\label{CFisUC}
    \item IMCCE, UMR8028 CNRS, Observatoire de Paris, PSL Universit\'e, 77 Av. Denfert-Rochereau, 75014 Paris, France \label{IMCCE}
    \item Department of Astrophysical Sciences, Princeton University, 4 Ivy Lane, Princeton, NJ 08544\label{Princ}
    \item Space sciences, Technologies and Astrophysics Research (STAR) Institute, Université de Liège, Allée du 6 Août 19C, 4000 Liège, Belgium\label{uliege2}
    \item SRON, Netherlands Institute for Space Research, Niels Bohrweg 4, NL-2333 CA, Leiden, The Netherlands\label{SRON}
    \item NASA Goddard Space Flight Center, Greenbelt, MD, USA\label{NASAGo}
    \item Integrated Space Science and Technology Institute, Department of Physics, American University, Washington DC, USA \label{AmUn}
    \item Department of Astronomy and Carl Sagan Institute, Cornell University, Ithaca, New York 14853, USA\label{Corn}
    \item Lunar \& Planetary Laboratory/Department of Planetary Sciences, University of Arizona, Tucson, AZ, USA \label{LPL}
    \item Blue Marble Space Institute of Science, 600 1st Avenue, 1st Floor, Seattle, Washington 98104, USA\label{BlueM}
    \item Institute of Planetary Research, German Aerospace Center (DLR), Rutherfordstrasse 2, D-12489 Berlin, Germany \label{DLR}
    \item Lockheed Martin Solar and Astrophysics Lab (LMSAL), Palo Alto, CA 94304, USA \label{Lockh}
    \item Department of the Geophysical Sciences, University of Chicago, 5734 S Ellis Ave., Chicago, IL 60637 \label{UChi}
    \item Center for Space and Habitability, University of Bern, Gesellschaftsstrasse 6, 3012 Bern, Switzerland\label{UBern}
    \item naXys Research Institute, Department of Mathematics, University of Namur, 61 Rue de Bruxelles, Namur, Belgium\label{UNam}
    \item Department of Aerospace, Physics and Space Sciences, Florida Institute of Technology, Melbourne, FL 32901\label{FIT}
    \item Institute for Astronomy, School of Physics, Zhejiang University, Hangzhou 310027, China\label{UZhej}
    \item Space Telescope Science Institute, 3700 San Martin Drive, Baltimore, MD 21218-2410, USA\label{STScI}
    \item Department of Astrophysical \& Planetary Sciences, University of Colorado Boulder, Boulder, CO, USA\label{UCol}
    \item Instituto de Astrofísica de Andalucía (IAA-CSIC), Glorieta de la Astronomía s/n, 18008 Granada, Spain\label{IAA}
    \item DTU Space, National Space Institute, Technical University of Denmark, Elektrovej 328, DK-2800 Kgs. Lyngby, Denmark\label{DTU}
    \item School of Physics \& Astronomy, University of Birmingham, Edgbaston, Birmingham, B15 2TT, UK \label{UBirm}
    \item NASA Goddard Institute for Space Studies, 2880 Broadway, New York, NY USA\label{NASAGISS}
    \item Department of Physics and Astronomy, Theoretical Astrophysics, Uppsala University, Uppsala, Sweden\label{UUpp}
\end{footnotesize}
\end{affiliations}

\begin{abstract}

Ultra-cool dwarf stars are abundant, long-lived, and uniquely suited to enable the atmospheric study of transiting terrestrial companions with JWST. Amongst them, the most prominent is the M8.5V star TRAPPIST-1 and its seven planets. While JWST Cycle 1 observations have started to yield preliminary insights into the planets, they have also revealed that their atmospheric exploration requires a better understanding of their host star. Here, we propose a roadmap to characterize the TRAPPIST-1 system---and others like it---in an efficient and robust manner with JWST. We notably recommend that---although more challenging to schedule---multi-transit windows be prioritized to mitigate the effects of stellar activity and gather up to twice more transits per JWST hour spent. We conclude that, for such systems, planets cannot be studied in isolation by small programs, but rather need large-scale, jointly space- and ground-based initiatives to fully exploit the capabilities of JWST for the exploration of terrestrial planets.


\end{abstract}

\flushbottom


Terrestrial planets surrounding M dwarfs are abundant \citep{Dressing2015,Gaidos2016,Ment2023}. As these stars dominate the galactic population \citep{Bochanski2010}, planetary systems around M dwarfs can be viewed as windows on the galactic terrestrial-planet population (Figure~1.a.). Unfortunately, terrestrial exoplanets are only amenable for atmospheric studies with JWST when found around mid to late M dwarfs (Figure~1.b.), specifically around ultra-cool dwarf stars (UCDs, with effective temperature less than 3000\,K) for those within the temperate zone---defined as receiving a flux between $4\times$ and $0.25\times$ that of the Earth\citep{Triaud2023}. Figure 1.b. highlights the handful of systems with transiting terrestrial exoplanets amenable for atmospheric characterization with JWST known to date (incl., L~98-59 \citep{Kostov2019}, LHS~1140 \citep{Dittmann2017}, LHS~3844 \citep{Vanderspek2019}, LP~791-18 \citep{Crossfield2019,Peterson2023}, LP~890-9\citep{Delrez2022}, TOI-540 \citep{Ment2021}, and TRAPPIST-1 \citep{Gillon2016,Gillon2017}). The small sizes of UCDs yield favorable planet-to-star radius ratios that enable the detectability of terrestrial planet atmospheres via transmission spectroscopy.  

If temperate planets around UCDs are able to acquire a moderate amount of atmospheric volatiles during planet formation\citep{Lichtenberg2022}, and preserve it during their star’s extended pre-main-sequence phase\citep{Baraffe1998,Baraffe2015}, our galaxy may be host to billions of habitable oases. Alternative scenarios, depending on volatile inventories and age, may include scorched desert, runaway greenhouse, or frigid, ice-locked ocean worlds\citep{Kane2014,Tian2015,Kane2019,Lichtenberg2019,Venturini2020,Way2020,Kimura2022}. Other factors such as large XUV fluxes, stellar winds, flares, and coronal mass ejections can evaporate and erode planetary atmospheres and remain at high levels on Gyr timescales,  thus supporting the desert fate \citep{Luger2015,Dong2017,lincowski2018,Dong2018}. In addition, after the protoplanetary-disk phase collisions with planetesimals may deliver volatiles to the outermost planets but erode the atmospheres of the innermost ones\citep{Kral2018}. As the observational constraints on both the effectiveness of these processes and the planets' original volatile reservoirs are limited, reliable predictions are poorly constraining. In some instances even H$_2$-dominated atmospheres can be sustained by a balance between outgassing and escape\citep{Hu2023}.  The presence of an atmosphere on terrestrial UCD planets must therefore be established empirically, an endeavor for which JWST is uniquely suited.

\section{TRAPPIST-1 as an opportunity and a test case.}

Understanding how planets form, assemble, and evolve around UCDs is a fundamental question of planet formation, and the TRAPPIST-1 system provides a benchmark for such studies\citep{Ormel2017}. 
Being able to study how the presence or absence of an atmosphere or how the properties of such an atmosphere vary over seven planetary configurations within a single system will be considerably more informative than a similar number of configurations in distinct systems with different histories and properties (incl. stellar activity). Similarly, comparative studies leveraging planets within the same system are key to enable habitability assessment\citep{Triaud2023}.

While multiplanetary systems are expected\citep{Weiss2018,Sandford2021,Mishra2021,Millholland2021,Goyal2022}, only the TRAPPIST-1 system has been identified with three or more transiting terrestrial planets and at least two temperate ones---despite dedicated surveys such as SPECULOOS \citep{Delrez2018,Burdanov2018} EDEN \citep{Gibbs2020}, and PINES\citep{Tamburo2022} searching for such systems around 20 times more UCDs than the original TRAPPIST survey \citep{Gillon2013a}. TRAPPIST-1 can thus be seen as both a unique opportunity and test case.

The radii of the seven TRAPPIST-1 planets have been measured to a precision of a few \% \citep{Ducrot2020}. Their masses have been constrained to the same precision via transit-timing variations.  The planets adhere to a single rocky mass--radius relation that can notably correspond to an iron depletion relative to Earth\citep{Agol2021}, or an Earth-like composition enhanced in volatiles\citep{Lichtenberg2022,Lichtenberg2019,Venturini2020}  possibly splitting the climatic evolution of the planets in- and outside of the long-lasting runaway greenhouse irradiation limit\citep{Dorn2021}. 

Regarding the planetary atmospheres, pilot surveys with transmission spectroscopy from HST/WFC3 and the mass--radius relationship of the planets have independently ruled out the presence of large hydrogen-dominated atmospheres for all the planets \citep{dewit2016,deWit2018,Wakeford2019,Turbet2020,Agol2021,Garcia2022,Gressier2022}. The next step in the characterization of these planets is thus to assess the presence of secondary cloud-free atmospheres, and---if present---to plan their detailed study. This opportunity has already motivated eleven JWST Cycle programs totaling over 400 JWST hr, corresponding to over 250hr of science time. Onwards, all reported time requirements refer to science time.

\section{Lessons from Cycle 1.}

As the first insights into the system gained from Cycle 1 observations are becoming public, we identify key lessons that may help guide future programs targeting terrestrial exoplanetary systems.

\subsection{Planets.}

To this date, only observations relative to the inner planets of the system have been made public: in emission\citep{Greene2023} and in transmission\citep{Lim2023} for planet b and in emission only for planet c\citep{Zieba2023}. The transmission spectrum of planet b appears dominated by stellar contamination while its emission reveals a very hot dayside, consistent with zero heat redistribution. On the other hand, planet c has a 15 microns brightness temperature that is less consistent with a null albedo bare rock, preliminary seen as indicative of a thin atmosphere or a higher surface albedo. Subsequent analyses however found a broad range of atmospheric configurations that cannot be ruled out by the data\citep{Lincowski2023}, incl. thick O$_2$-dominated atmospheres.

\subsection{Star.}

JWST Cycle 1 programs targeting transits of planets around K- and M-type stars including TRAPPIST-1 have shown that a key limitation stems from the effects of stellar activity in the time domain (e.g., flares\citep{Howard2023}) and in the wavelength domain (e.g., stellar contamination) \citep{Moran2023,Lim2023} (Figure 2). 


 Cycle 1 observations of TRAPPIST-1 have shown that flare events occur during most transit observations and have intensities up to several thousands of ppm in the near infrared (Figure 2.a.). 
Flares constitute a significant time- and wavelength-dependent signal that contaminate transit depth measurements by affecting the out-of-transit baseline and/or the in-transit data. A mini-flare can also be mistaken for a spot-crossing event as was observed on July 20$^{\rm th}$ 2022, just before the egress of TRAPPIST-1~b\citep{Lim2023}. 

In addition, the chromatic transit depth, or ``transmission spectrum,'' of a planet contains only information related to the wavelength-dependent opacity of a planet's atmosphere if its star is a limb-darkened but otherwise featureless disk. As most stars do show surface features in the form of spots and faculae, the transmission spectrum also contains a stellar contribution due to the difference between the hemisphere-averaged emission spectrum of the star and the transit-chord-averaged one (Figure 2.d.), a phenomenon known as the ``transit light source'' (TLS) effect\citep{rackham2018,Rackham2019A}. Furthermore, spectra of magnetic features are not well described by 1D models, possibly leading to biased interpretations\citep{Witzke2021,Rustamkulov2023}.

Prior to JWST, studies already highlighted the effect of stellar contamination for the TRAPPIST-1 system \citep{Zhang2018,Wakeford2019,Ducrot2020,Garcia2022}. Space- and ground-based data showed that stellar contamination in the system perturbs the apparent transit depth of each planet by up to 10\% (Figure~2.d.), resulting in spectral signals of up to $\sim$700 p.p.m. in amplitude, larger than the $\sim$200-p.p.m. signals expected from secondary atmospheres (Figures~1.b. and 2.d.). The first JWST observations of the system confirm that stellar contamination can be as high as 600\,ppm, although it is not consistent from visit to visit and planet to planet\citep{Lim2023}.



\section{Towards a roadmap for JWST's exploration of TRAPPIST-1-like systems}

Detecting the $\sim$200-p.p.m. signal of a secondary atmosphere in the TRAPPIST-1 system will necessitate anywhere between a few to several dozen transits depending on the planet and its bulk abundances \citep{Morley2017,Krissansen2018,Lustig-Yaeger2019,Fauchez2019,Wunderlich2019,Gialluca2021}. This corresponds to upwards of 46 transits for the seven planets of the TRAPPIST-1 system. Once identified, the in-depth characterization of a secondary atmosphere would require a similar level of commitment\citep{Triaud2023}. Thus, the detailed atmospheric study of terrestrial planet atmospheres with JWST will be a multi-step process likely requiring hundreds of transit or occultation observations spreading over $>$5 cycles.


These estimates assume that the stellar effects mentioned above can be fully mitigated. To this end, the continuum jitter associated with flare activity must be corrected and so must the TLS effect. This is an inescapable prerequisite to reliably extracting and interpreting $\leq$200-ppm atmospheric signals expected for this system. This situation is akin to the precision radial velocity field that, about one decade ago, had to learn how to extract Keplerian signals significantly smaller than those generated by stellar activity. Thanks to various activity indicators and novel analysis methods, Keplerian signals smaller than those generated by stellar activity are now routinely detected close to the photon-noise limit (e.g., as for Proxima d \citep{Faria2022}). 


Flares have a $\sim$30\% chance of significantly biasing the out-of-transit baseline of a standard 6-hr transit observation of TRAPPIST-1 (e.g., Figure 2.a.) as they occur at a rate of 3.6$\pm^{2.1}_{1.3}$ flare/day and last between thirty minutes and three hours\citep{Howard2023}. While extending baselines when performing transit observations is a good first-step mitigation strategy, it is key to derive a better understanding of a UCD's flares via the monitoring of its activity over a long period of time. Doing so will yield empirical calibrations between H$\alpha$, recombination lines, and the variable, wavelength-dependent continuum flux. Indeed, flares are detected in H$\alpha$ along with several other recombination lines of the Paschen and Brackett series (Figure~2.c.). A full stellar rotation curve ($\sim$80\,hr) will yield a sufficient sample ($N\geq10$) to inform these relationships. 


Correcting for the TLS effect utilizes the out-of-transit stellar spectrum to derive the temperatures and covering fractions of different components to correct for their contributions to the stellar contamination \citep{Zhang2018,Wakeford2019,Garcia2022}. Doing so requires reliable stellar spectral models to break otherwise-limiting degeneracies\citep{Rackham2023}. Figure~2.e-f. presents the out-of-transit JWST/NIRISS spectra of the G8V star WASP-39 and of the M8.5V star TRAPPIST-1 with their respective best fit, highlighting the significant model inaccuracy for TRAPPIST-1, which are primarily due to stellar models being streched beyond their intended usage to approximate spots and faculae. 

Modelling of cool stars relies on a number of crucial steps, including accurate treatment of molecular opacity and the equation of state. Fortunately, tight constraints on the emission spectra of heterogeneities can also be derived empirically from a full stellar rotation via time-resolved spectroscopy\citep{Berardo2024}, thereby providing a unique opportunity to benchmark a new generation of stellar models (e.g., 3D MHD code MURaM \citep{Vogler2005} and MPS-ATLAS spectra \citep{Witzke2021}). In the meantime, we advocate to perform first-order corrections of the TLS effect by leveraging multi-transit windows with at least one planet expected to have a marginal atmospheric signature in transmission, thereby constraining the TLS signal at that epoch.


Constraints on the timescales of temporal variability of the covering fractions will in turn provide constraints on whether the variability arises from rotational modulation of active regions, physical evolution of active regions, or both. Two puzzles of TRAPPIST-1's variability \citep{Morris2018} could then be explored: (1)\ the existence of bright spots of line emission producing the $\sim$1\% variability seen in the Kepler band but not in Spitzer’s, and (2) the coincidence of flares and the steepest rise in the spot flux observed with Kepler.

With the previous considerations in mind, we propose the following roadmap for the exploration of a terrestrial planet system with JWST (flowchart in Figure 3), which we expand upon below:

\begin{enumerate}
    \item Gather MIRI emission observations of the inner planets to assess the presence of an atmosphere via $\sim$10 eclipses ($\leq$50hr of science time) per planet (e.g., PIDs 1177, 1279, and 2304, with Greene, Lagage, and Kreidberg as PI, respectively) and/or a joint phase curve ( $\leq$60\,hr, e.g. PID 3077 with Gillon as PI).  ;
    \item If one of the inner planets does have a ``featureless'' atmosphere such that its transmission spectrum primarily records the TLS effect at that epoch, it can be used to extract the fist order correction for TLS for the atmospheric exploration of other planets ($\leq$10 NIRSpec/PRISM transits per planets for a total of $\leq$200\,hr for the system via multi-transit windows);
    \item If all inner planets appear to have an atmosphere, it will only be possible to correct for TLS via empirical constraints on the stellar models. Therefore, a full NIRISS/SOSS stellar rotation curve\citep{Berardo2024} would be needed to search for the presence of atmosphere around the other planets ($\sim$80\,hr in addition to the $\sim$200\,hr mentioned at point 2). 
    \item If an atmosphere is detected, its in-depth characterization may take upwards of $\sim$300\,hr\citep{Triaud2023} and will require the empirical emission spectra of the stellar heterogeneities mentioned at point 3 to ensure a thorough correction of TLS.
     
\end{enumerate}

MIRI emission observations of the inner planets should be prioritized at the beginning of the TRAPPIST-1 roadmap\citep{Greene2023,Zieba2023}. These measurements are not sensitive to stellar heterogeneity, and the results will provide important context for the other planets in the system. If an atmosphere has survived at the relatively high irradiation incident on planet b or c, that would be an encouraging prospect that the cooler planets have atmospheres too. However, an absence of atmosphere around the innermost planets has limited to no implications on the odds of outer planets retaining substantial atmospheres\citep{Krissansen2023}. The inner planets are both the least likely to have retained their atmospheres and the ones with the shortest period (i.e., with the greatest probability of having consecutive transits) and thus may offer a valuable opportunity to support preliminary correction of TLS when gathering multi-transit observations (Figure 4). Such preliminary corrections would be sufficiently precise (to within $\leq$50ppm, Figure 4.d) to support revealing secondary atmospheres around other planets, but would not support a later in-depth exploration which would require stellar models of sufficient fidelity. The TLS associated with an atmosphereless planet is not identical to the one recorded by other planets in quasi-contemporaneous transits due to their different timings, impact parameters, and sizes. While most MIRI emission observations have been carried out in Imaging, applications with LRS in fixed-slit mode are expected to yield a substantial ($\sim$2.5) SNR increase (e.g., PID 6219, PI Dyrek).  An extensive MIRI emission survey of the TRAPPIST-1 planets could be executed as early as Cycle 3 as part of the exoplanet 500-hour Director Discretionary Time (DDT) program recommended by the exoplanet working group to focus solely on secondary eclipse observations \citep{Redfield2024}.


To assess the presence of atmosphere amenable to in-depth characterization with JWST around the other TRAPPIST-1 planets, we recommend leveraging multi-transit windows as successfully implemented during the reconnaissance of the system with HST (PIDs~14500\&14873, PI de Wit) for the following three reasons (highlighted in Figure 4). First, multi-transit windows decrease the overhead per transit and can yield up to twice more transits per JWST hour spent, thereby allowing us to save hundreds of JWST hours on this system alone. As $\sim$46\% of TRAPPIST-1's transit events happens within 5 hours of another transit (Figure~5.a.)--which is the typical duration for transit observations of the system--the resulting increase in scheduling constraint can be substantially mitigated. Second, multi-transit windows provide an extended baseline, thereby mitigating the relative effect of flare events. Third, back-to-back transits provide a quasi-contemporaneous scan of the star, increasing substantially the information content of the dataset and supporting notably better constraints on the stellar contamination (Figure 4.d.). Consecutive transits of adjacent chords will sweep up to 60\% of the visible stellar hemisphere, facilitating searches for active regions from the stellar equator to stellar latitudes of $35^{\circ}$ \citep{Delrez2018,Agol2021} and helping map the star \citep{Luger2021a,Luger2021b,Luger2022}, thereby informing the possible evolution of atmospheres and biospheres.  Finally, stellar scans will help constrain their mutual inclinations of the planets (i.e., on which side of the ecliptic each planet is) by comparing their in-transit spot-crossing patterns (Figure 4.e.). This approach will be used for the first time to assess the presence of an atmosphere around TRAPPIST-1~e, through the observations of 15 quasi-contemporaneous transits (80hr of science time) with planet b spread over Cycles 3 \& 4 (PID 6456, PI Allen).

Following the approved Cycle 1 programs and Ref.\citep{Lustig-Yaeger2019}, we find that up to sixteen of such multi-transit windows are needed to complement the Cycle 1 programs and reliably assess the presence of atmospheres amenable for further exploration around each of the TRAPPIST-1 planets. The exact number depends on the series of transits gathered in each window. These windows typically spread over two cycles due to their scheduling constraints and require $\sim$100\,hr of science time (Figure 5.b.). NIRSpec/PRISM is best suited for this search as it covers a wide wavelength coverage suitable for TLS correction and atmospheric-component identification\citep{Lustig-Yaeger2019,Triaud2023}.  


If signs of a planetary atmosphere are found, high-fidelity stellar models will be required to support its further study. To this end, we recommend acquiring a stellar rotation curve to derive the empirical emission spectra of surface heterogeneities and correct for the TLS effect\citep{Berardo2024}, as well as empirical calibrations to support the corrections for flares (Section 3.1.). We recommend acquiring the stellar rotation curve with a maximum number of contemporaneous transits. We have identified at least four windows per Cycle offering nine bonus planetary transits during a full rotation (Figure~4.a.). NIRISS/SOSS is the optimal set up for this task owing to its spectral coverage and resolving power, allowing to constrain key molecular features and spectral lines while preventing full saturation. Joint space- and ground-based observations covering a broad wavelength range are highly recommended---HST for UV/VIS monitoring in particular.

In addition to the benefits of joint (i.e., simultaneous) observations, long-term parallel monitoring from the ground can complement space-based monitoring, providing independent constraints on stellar activity. Joint physical modelling of photometry and activity indicators from high-precision spectroscopy can reconstruct the surface distribution of active regions on the face of the star and their time evolution \citep{Mallonn2018}, therefore providing other means to correct for stellar contamination during transits \citep{Rosich2020}. Multi-technique and  multi-band observations, covering the widest possible wavelength range, make it possible to disentangle most of the parameter degeneracies of stellar active regions
\citep{Perger2023} and render their mapping possible. We thus recommend to bracket JWST exoplanet transit observations with ground-based monitoring to ensure that the best possible understanding of the stellar surface heterogeneities at the relevant epochs are attained in order to reliably disentangle between the stellar and planetary signals in transit. 

While the in-depth atmospheric exploration of a terrestrial planet will require high-fidelity stellar models to support the sufficient mitigation of stellar-activity effects, we recommend not delaying the acquisition of the necessary transit observations---once an atmosphere amenable for such study has been detected. Indeed, the number of transits required for a detailed atmospheric study will range from $\sim40$ to $\geq$100 (e.g., for habitability and inhabitability assessments\citep{Triaud2023}). For planets with orbital periods ranging from a few to tens of days, such a requirement will translate into observational programs spreading over 2 to $\geq$10 years.

As the steps towards an efficient characterization of terrestrial-planet systems with JWST will require tight scheduling constraints, a significant time commitment, and a wide range of expertise and facilities, we argue that a single, coherent, cross-disciplinary observational program is needed per system of interest.

\vspace{0.5cm}

\newpage
\pagebreak
\clearpage

\section{Figures}

\begin{figure*}[h!]
\centering
\hspace*{-20mm}\includegraphics[trim={0cm 0cm 0cm 0cm},clip, angle=0, width=1.25\textwidth ]{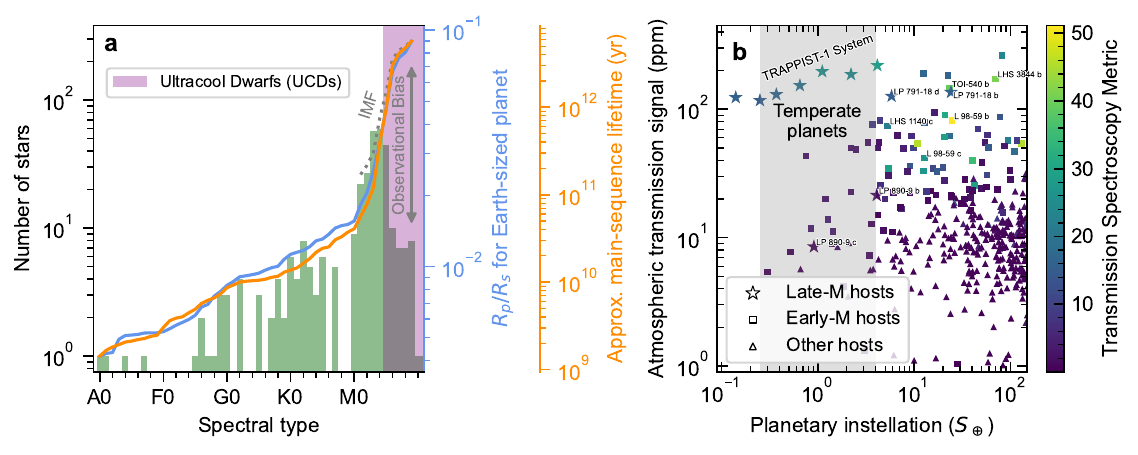}
\caption{\textbf{The Ultra-Cool Dwarf Stars Opportunity.} \textbf{a.} Histogram of spectral types of all stars within 10\,pc observed by Gaia\citep{Reyle2021} together with the main-sequence lifetime (orange)\citep{Hansen1994} and expected transit depth for a terrestrial planet for each type (blue).
The stellar initial mass function\citep{Kroupa2001} is shown as a dotted gray line.
The region of UCDs is shaded in purple.
\textbf{b.} Atmospheric signal amplitude and transmission spectroscopy metric\citep{Kempton2018} for known terrestrial exoplanets. 
Marker sizes are proportional to planet sizes, and planets transiting late-M, early-M, and other hosts are shown as stars, squares, and triangles, respectively. 
The region of temperate planets with instellations between 0.25 and 4\,$S_\oplus$ is highlighted.
\label{fig:figure1}}
\end{figure*}

\newpage

\begin{figure*}[h!]
\centering
\vspace*{-30mm}{\includegraphics[trim={0cm .3cm 0cm 0cm},clip, angle=0, width=0.8\textwidth, height=0.7\textheight, keepaspectratio]{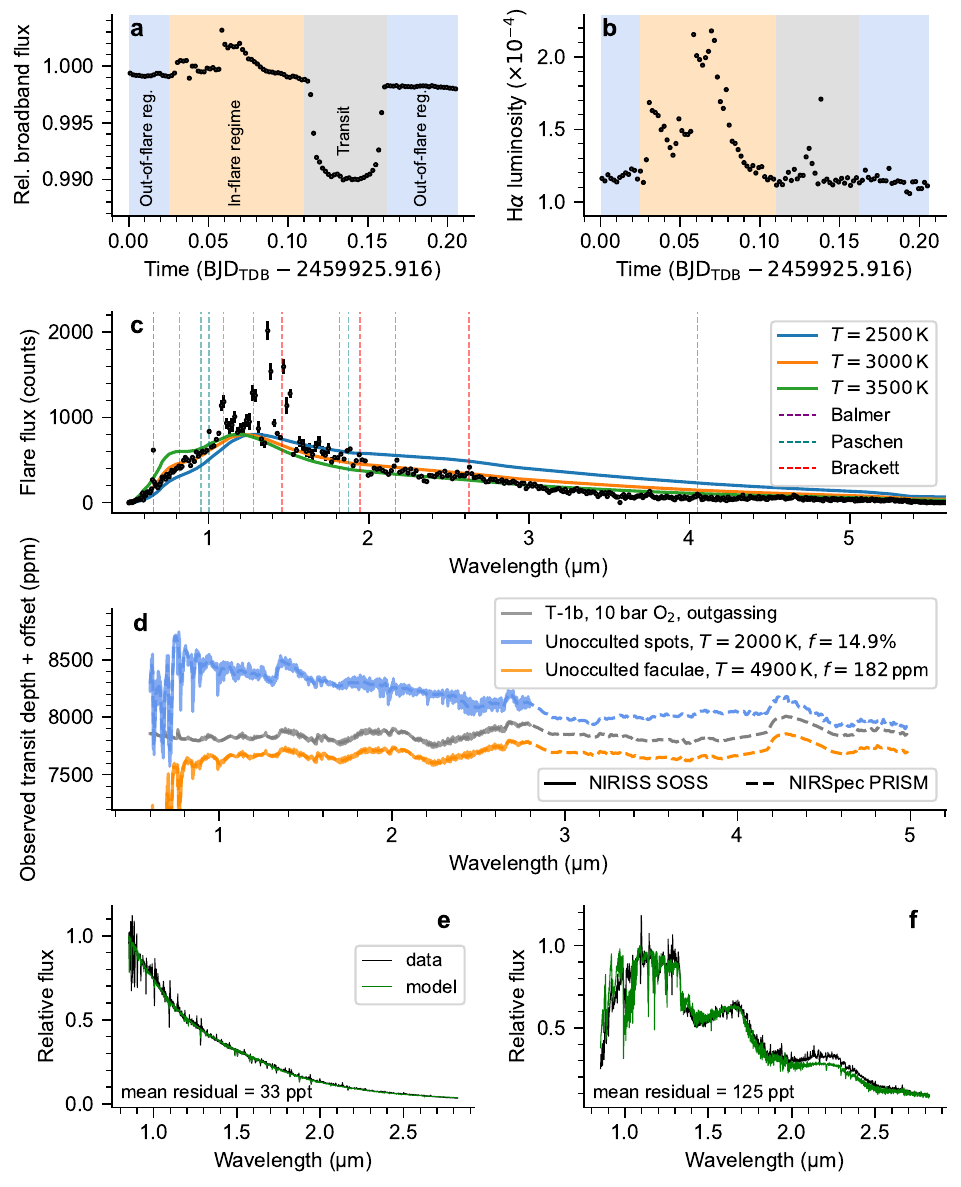}}
\caption{\textbf{Stellar activity challenges spectral analysis of M-dwarf systems.} 
\textbf{a.} White-light NIRSpec/PRISM light-curve of TRAPPIST-1~g transit showing multiple flares and a planetary transit from Ref.\citep{Howard2023}. 
\textbf{b.} Integrated flux in the H$\alpha$ stellar line during this observation. 
\textbf{c.} Extracted emission spectrum of the flare (black points), consistent with a 3000\,K blackbody. 
Models for 2500\,K, 3000\,K, and 3500\,K blackbodies are shown in blue, orange, and green, respectively.
Purple, teal, and red dashed lines indicate locations of Balmer, Paschen, and Brackett recombination lines, respectively. 
\textbf{d.} Model transmission spectra of TRAPPIST-1\,b for an O$_2$-dominated atmosphere in the case of no stellar contamination (grey), unocculted spots (blue), and unocculted faculae (orange).
Spot and facula parameters are drawn from ref.\,\citep{Garcia2022}.
\textbf{e.-f.} NIRISS/SOSS spectra (black) with best-fit stellar models (green) for the G-dwarf WASP-39 and M-dwarf TRAPPIST-1.
\label{fig:figure2}}
\end{figure*}

\begin{figure*}[h!]
\hspace{-2cm}{\includegraphics[trim={0cm 18cm 0cm 0cm},clip, angle=0, width=1.3\textwidth ]{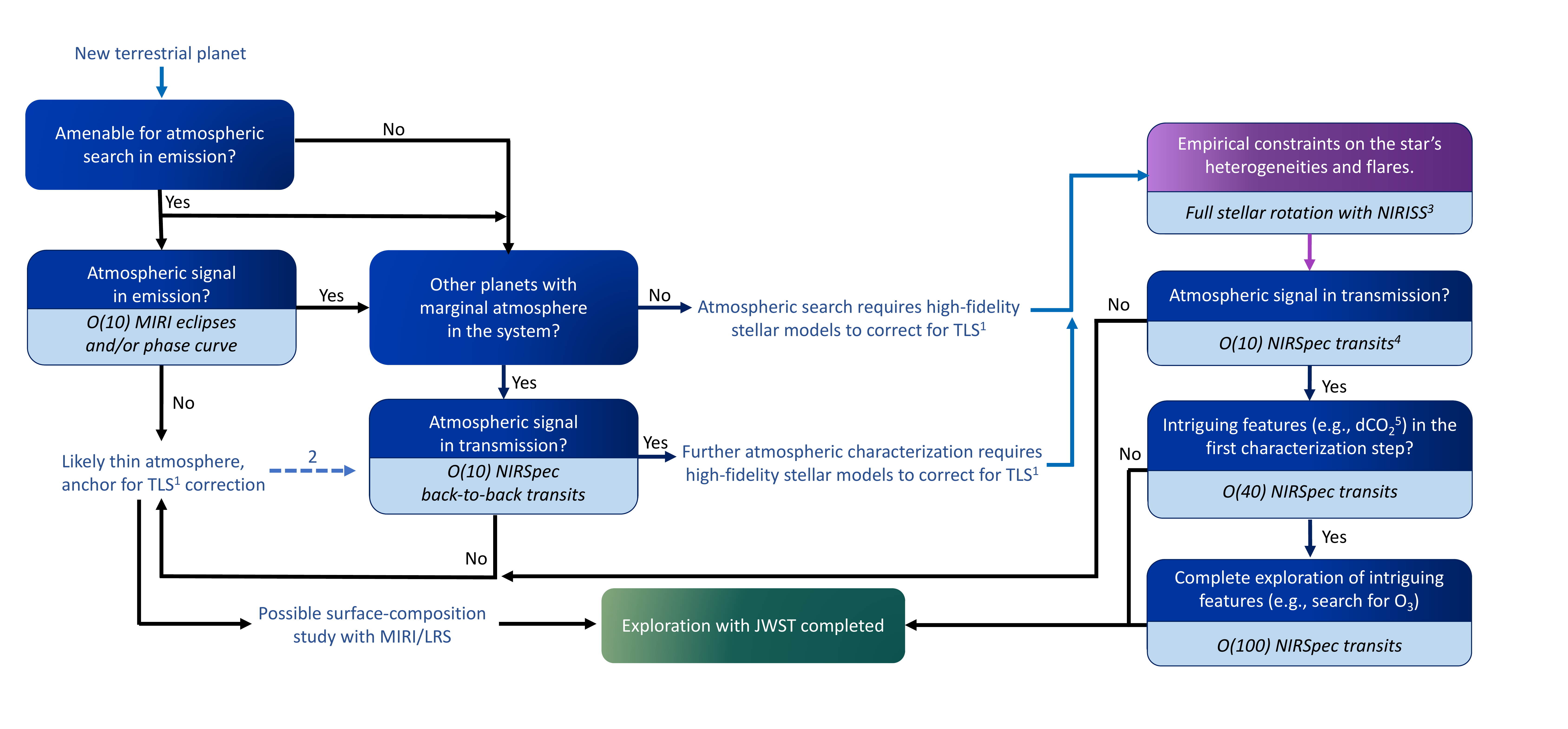}}
\caption{\textbf{Flowchart of a roadmap for the atmospheric characterization of terrestrial exoplanets with JWST.} Dark blue boxes present key questions along the roadmap for characterization with JWST. Light blue boxes present the observations needed to support answering specific questions. The purple box present complementary (not planet related) information required to support further steps on the roadmap. The green box represent the final step on the roadmap, i.e. a completed exploration. $^1$ TLS (transit light source effect\citep{rackham2018}) refers here to stellar activity/contamination. $^2$ The dashed arrow represents the fact that an “airless” planet can later be used to correct for TLS via back-to-back transits. $^3$ A full rotation light-curve\citep{Berardo2024} should be observed over a window that maximizes the number of transits, simultaneously with other facilities that cover complementary wavelength ranges. $^4$ No additional measurements needed if atmospheric signal detected at a previous step. $^5$ dCO$_2$ refers to atmospheric carbon depletion\citep{Triaud2023}.
\label{fig:figure3}}
\end{figure*}

\begin{figure*}[h!]
\vspace{-2cm}{\includegraphics[trim={0cm 0cm 0cm 0cm},clip, angle=0, width=0.9\textwidth ]{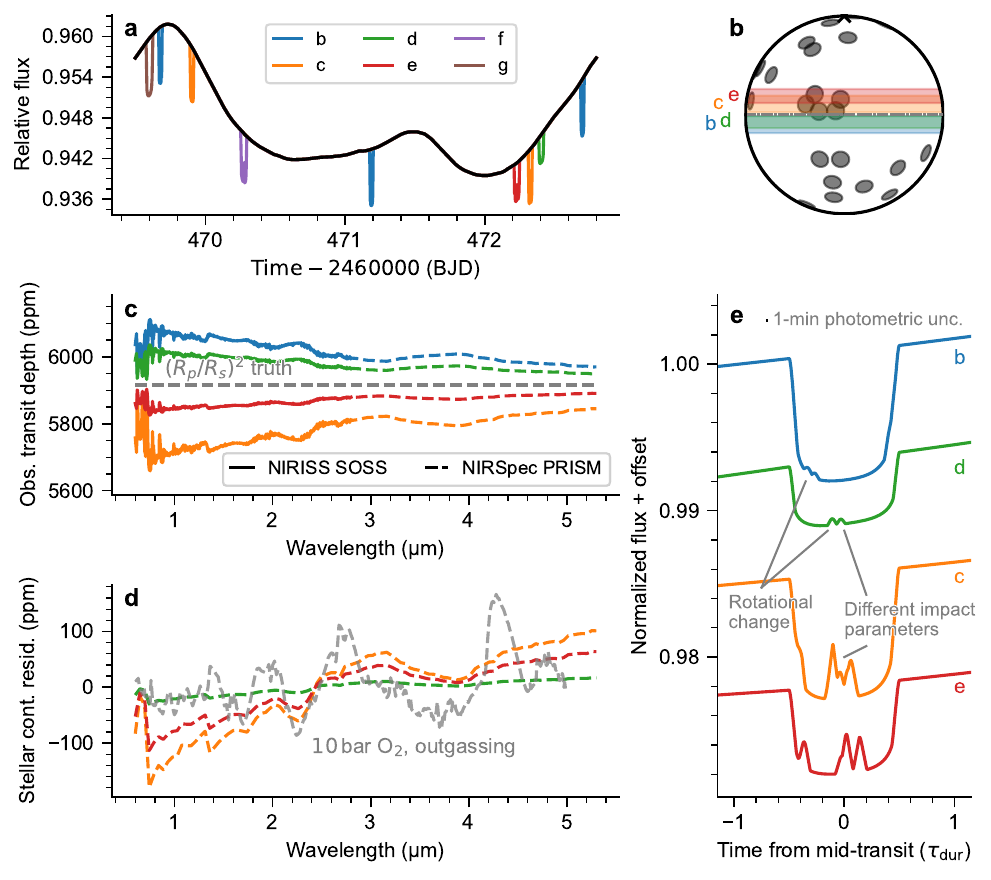}}
\caption{\textbf{Constraining the surface heterogeneities of a host star.}
\textbf{a.} A simulated 3.3-d full rotation curve of TRAPPIST-1 (black line) featuring nine transits of six planets (coloured lines).
\textbf{b.} A snapshot of the spot distribution at the last rotational phase of the simulation shown in panel a, with the transit chords of planets b, c, d, and e.
\textbf{c.} Contaminated transmission spectra for the same transit train of planets b--e (coloured lines).
A nominal planet radius of 1\,$R_\oplus$ (dashed gray line) is used for each planet to ease comparisons of the TLS effect for each planet.
\textbf{d.} The residual stellar contamination signals for the simulated transits of planets c--e (coloured lines), after correcting for the TLS effect using the in-transit signal from planet~b. 
A median-subtracted model of an atmosphere (see Figure 2) is shown for comparison (gray dashed line).
\textbf{e.} Close-up view of the transit profiles for the four-planet transit train, highlighting repeated occultations of the same active regions by planets b and d as well as c and e, whose transit chords overlap in this simulation.
\label{fig:figure4}}
\end{figure*}

\begin{figure*}[h!]
\hspace{-2cm}\includegraphics[trim={0cm 10cm 0cm 0cm},clip, angle=0, width=1.75\textwidth ]{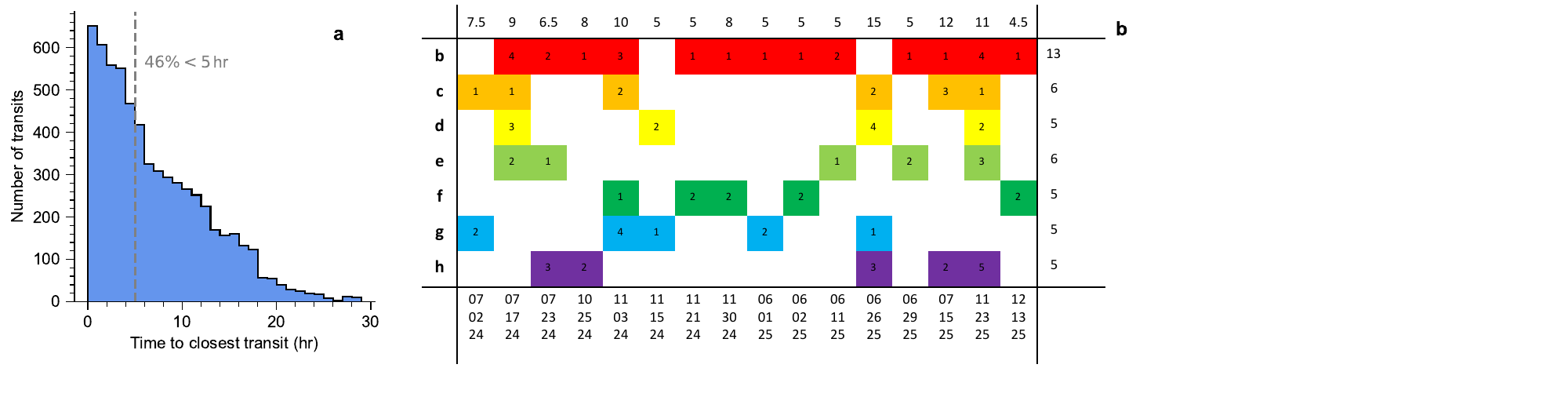}
\caption{\textbf{On the frequency of planetary transits in the TRAPPIST-1 system.} 
\textbf{a.} Histogram of time-to-closest-transit for each TRAPPIST-1 transit event between 2015--2025\citep{Agol2021} 
\textbf{b.} Example of a suite of sixteen observation windows to complete the roadmap step \# 2 (atmospheric assessment for all seven TRAPPIST-1 planets) by 01/01/26. The date and duration of each window are reported in the bottom and top rows, respectively. The total number of transits to be observed with this suite of observations is reported in the last column (e.g., 13 for planet b).  Numbers in colored boxes indicate the order in which the transits occur within each window.
\label{fig:figure5}}
\end{figure*}

\newpage
\pagebreak
\clearpage

\begin{addendum}
 \item 
 
\item[Author Contributions] 
J.d.W. and R.D. led this community-supported project. B.V.R. led the production of the figures together with J.d.W., R.D., O.L., B.B., P.N., D.B., and Z.d.B.. E.D. and L.K. led the discussions associated with emission studies. I.R. led the discussions associated with complementary ground-based studies. A.I., A.S., N.K., and V.W. led the discussions related to stellar models. All authors discussed the topics in the paper, contributed to the writing and commented on the manuscript at all stages.

\item[Competing Interests] The authors declare no competing interests.

\item[Data Availability] 
Data presented in Figure~1.b.\ are drawn from the NASA Exoplanet Archive (Accessed 29\,Jun\,2023). The transit timings behind the histogram in Figure 5.a.\ and the suite of selected multi-transit windows in Figure 5.b.\ are obtained from Ref.\citep{Agol2021}.
All data used to create the figures are publicly available at \url{https://zenodo.org/records/11388689}.

\item[Code Availability] The stellar spectra (Figures 2 \& 4) were generated with \texttt{speclib} \citep{speclib}. The atmospheric spectra were generated with \texttt{Tierra}\citep{Niraula2022}. The light curve model (Figure 4) was generated with \texttt{fleck}\citep{Morris2020}.

 \item[Correspondence] Correspondence and requests for materials should be addressed to jdewit@mit.edu. 

\end{addendum}

\section*{References}

\begin{thebibliography}{10}
\expandafter\ifx\csname url\endcsname\relax
  \def\url#1{\texttt{#1}}\fi
\expandafter\ifx\csname urlprefix\endcsname\relax\def\urlprefix{URL }\fi
\providecommand{\bibinfo}[2]{#2}
\providecommand{\eprint}[2][]{\url{#2}}

\bibitem{Greene2023}
\bibinfo{author}{{Greene}, T.~P.} \emph{et~al.}
\newblock \bibinfo{title}{{Thermal emission from the Earth-sized exoplanet TRAPPIST-1 b using JWST}}.
\newblock \emph{\bibinfo{journal}{\nat}} \textbf{\bibinfo{volume}{618}}, \bibinfo{pages}{39--42} (\bibinfo{year}{2023}).
\newblock \eprint{2303.14849}.

\bibitem{Zieba2023}
\bibinfo{author}{{Zieba}, S.} \emph{et~al.}
\newblock \bibinfo{title}{{No thick carbon dioxide atmosphere on the rocky exoplanet TRAPPIST-1 c}}.
\newblock \emph{\bibinfo{journal}{arXiv e-prints}} \bibinfo{pages}{arXiv:2306.10150} (\bibinfo{year}{2023}).
\newblock \eprint{2306.10150}.

\bibitem{Lim2023}
\bibinfo{author}{{Lim}, O.} \emph{et~al.}
\newblock \bibinfo{title}{{Atmospheric Reconnaissance of TRAPPIST-1 b with JWST/NIRISS: Evidence for Strong Stellar Contamination in the Transmission Spectra}}.
\newblock \emph{\bibinfo{journal}{\apjl}} \textbf{\bibinfo{volume}{955}}, \bibinfo{pages}{L22} (\bibinfo{year}{2023}).
\newblock \eprint{2309.07047}.

\bibitem{dewit2016}
\bibinfo{author}{{de Wit}, J.} \emph{et~al.}
\newblock \bibinfo{title}{{A combined transmission spectrum of the Earth-sized exoplanets TRAPPIST-1 b and c}}.
\newblock \emph{\bibinfo{journal}{\nat}} \textbf{\bibinfo{volume}{537}}, \bibinfo{pages}{69--72} (\bibinfo{year}{2016}).
\newblock \eprint{1606.01103}.

\bibitem{deWit2018}
\bibinfo{author}{{de Wit}, J.} \emph{et~al.}
\newblock \bibinfo{title}{{Atmospheric reconnaissance of the habitable-zone Earth-sized planets orbiting TRAPPIST-1}}.
\newblock \emph{\bibinfo{journal}{Nature Astronomy}} \textbf{\bibinfo{volume}{2}}, \bibinfo{pages}{214--219} (\bibinfo{year}{2018}).
\newblock \eprint{1802.02250}.

\bibitem{Triaud2023}
\bibinfo{author}{{Triaud}, A.} \emph{et~al.}
\newblock \bibinfo{title}{{Atmospheric carbon depletion as a tracer of water oceans and biomass on temperate terrestrial exoplanets}}.
\newblock \emph{\bibinfo{journal}{arXiv e-prints}} \bibinfo{pages}{arXiv:2310.14987} (\bibinfo{year}{2023}).
\newblock \eprint{2310.14987}.

\bibitem{Dressing2015}
\bibinfo{author}{{Dressing}, C.~D.} \& \bibinfo{author}{{Charbonneau}, D.}
\newblock \bibinfo{title}{{The Occurrence of Potentially Habitable Planets Orbiting M Dwarfs Estimated from the Full Kepler Dataset and an Empirical Measurement of the Detection Sensitivity}}.
\newblock \emph{\bibinfo{journal}{\apj}} \textbf{\bibinfo{volume}{807}}, \bibinfo{pages}{45} (\bibinfo{year}{2015}).
\newblock \eprint{1501.01623}.

\bibitem{Gaidos2016}
\bibinfo{author}{{Gaidos}, E.}, \bibinfo{author}{{Mann}, A.~W.}, \bibinfo{author}{{Kraus}, A.~L.} \& \bibinfo{author}{{Ireland}, M.}
\newblock \bibinfo{title}{{They are small worlds after all: revised properties of Kepler M dwarf stars and their planets}}.
\newblock \emph{\bibinfo{journal}{\mnras}} \textbf{\bibinfo{volume}{457}}, \bibinfo{pages}{2877--2899} (\bibinfo{year}{2016}).
\newblock \eprint{1512.04437}.

\bibitem{Ment2023}
\bibinfo{author}{{Ment}, K.} \& \bibinfo{author}{{Charbonneau}, D.}
\newblock \bibinfo{title}{{The Occurrence Rate of Terrestrial Planets Orbiting Nearby Mid-to-late M Dwarfs from TESS Sectors 1-42}}.
\newblock \emph{\bibinfo{journal}{arXiv e-prints}} \bibinfo{pages}{arXiv:2302.04242} (\bibinfo{year}{2023}).
\newblock \eprint{2302.04242}.

\bibitem{Bochanski2010}
\bibinfo{author}{{Bochanski}, J.~J.} \emph{et~al.}
\newblock \bibinfo{title}{{The Luminosity and Mass Functions of Low-mass Stars in the Galactic Disk. II. The Field}}.
\newblock \emph{\bibinfo{journal}{\aj}} \textbf{\bibinfo{volume}{139}}, \bibinfo{pages}{2679--2699} (\bibinfo{year}{2010}).
\newblock \eprint{1004.4002}.

\bibitem{Lichtenberg2022}
\bibinfo{author}{{Lichtenberg}, T.} \& \bibinfo{author}{{Clement}, M.~S.}
\newblock \bibinfo{title}{{Reduced Late Bombardment on Rocky Exoplanets around M Dwarfs}}.
\newblock \emph{\bibinfo{journal}{\apjl}} \textbf{\bibinfo{volume}{938}}, \bibinfo{pages}{L3} (\bibinfo{year}{2022}).
\newblock \eprint{2209.14037}.

\bibitem{Baraffe1998}
\bibinfo{author}{{Baraffe}, I.}, \bibinfo{author}{{Chabrier}, G.}, \bibinfo{author}{{Allard}, F.} \& \bibinfo{author}{{Hauschildt}, P.~H.}
\newblock \bibinfo{title}{{Evolutionary models for solar metallicity low-mass stars: mass-magnitude relationships and color-magnitude diagrams}}.
\newblock \emph{\bibinfo{journal}{\aap}} \textbf{\bibinfo{volume}{337}}, \bibinfo{pages}{403--412} (\bibinfo{year}{1998}).
\newblock \eprint{astro-ph/9805009}.

\bibitem{Baraffe2015}
\bibinfo{author}{{Baraffe}, I.}, \bibinfo{author}{{Homeier}, D.}, \bibinfo{author}{{Allard}, F.} \& \bibinfo{author}{{Chabrier}, G.}
\newblock \bibinfo{title}{{New evolutionary models for pre-main sequence and main sequence low-mass stars down to the hydrogen-burning limit}}.
\newblock \emph{\bibinfo{journal}{\aap}} \textbf{\bibinfo{volume}{577}}, \bibinfo{pages}{A42} (\bibinfo{year}{2015}).
\newblock \eprint{1503.04107}.

\bibitem{Kane2014}
\bibinfo{author}{{Kane}, S.~R.}, \bibinfo{author}{{Kopparapu}, R.~K.} \& \bibinfo{author}{{Domagal-Goldman}, S.~D.}
\newblock \bibinfo{title}{{On the Frequency of Potential Venus Analogs from Kepler Data}}.
\newblock \emph{\bibinfo{journal}{\apjl}} \textbf{\bibinfo{volume}{794}}, \bibinfo{pages}{L5} (\bibinfo{year}{2014}).
\newblock \eprint{1409.2886}.

\bibitem{Tian2015}
\bibinfo{author}{{Tian}, F.} \& \bibinfo{author}{{Ida}, S.}
\newblock \bibinfo{title}{{Water contents of Earth-mass planets around M dwarfs}}.
\newblock \emph{\bibinfo{journal}{Nature Geoscience}} \textbf{\bibinfo{volume}{8}}, \bibinfo{pages}{177--180} (\bibinfo{year}{2015}).

\bibitem{Kane2019}
\bibinfo{author}{{Kane}, S.~R.} \emph{et~al.}
\newblock \bibinfo{title}{{Venus as a Laboratory for Exoplanetary Science}}.
\newblock \emph{\bibinfo{journal}{Journal of Geophysical Research (Planets)}} \textbf{\bibinfo{volume}{124}}, \bibinfo{pages}{2015--2028} (\bibinfo{year}{2019}).
\newblock \eprint{1908.02783}.

\bibitem{Lichtenberg2019}
\bibinfo{author}{{Lichtenberg}, T.} \emph{et~al.}
\newblock \bibinfo{title}{{A water budget dichotomy of rocky protoplanets from $^{26}$Al-heating}}.
\newblock \emph{\bibinfo{journal}{Nature Astronomy}} \textbf{\bibinfo{volume}{3}}, \bibinfo{pages}{307--313} (\bibinfo{year}{2019}).
\newblock \eprint{1902.04026}.

\bibitem{Venturini2020}
\bibinfo{author}{{Venturini}, J.}, \bibinfo{author}{{Guilera}, O.~M.}, \bibinfo{author}{{Haldemann}, J.}, \bibinfo{author}{{Ronco}, M.~P.} \& \bibinfo{author}{{Mordasini}, C.}
\newblock \bibinfo{title}{{The nature of the radius valley. Hints from formation and evolution models}}.
\newblock \emph{\bibinfo{journal}{\aap}} \textbf{\bibinfo{volume}{643}}, \bibinfo{pages}{L1} (\bibinfo{year}{2020}).
\newblock \eprint{2008.05513}.

\bibitem{Way2020}
\bibinfo{author}{{Way}, M.~J.} \& \bibinfo{author}{{Del Genio}, A.~D.}
\newblock \bibinfo{title}{{Venusian Habitable Climate Scenarios: Modeling Venus Through Time and Applications to Slowly Rotating Venus-Like Exoplanets}}.
\newblock \emph{\bibinfo{journal}{Journal of Geophysical Research (Planets)}} \textbf{\bibinfo{volume}{125}}, \bibinfo{pages}{e06276} (\bibinfo{year}{2020}).
\newblock \eprint{2003.05704}.

\bibitem{Kimura2022}
\bibinfo{author}{{Kimura}, T.} \& \bibinfo{author}{{Ikoma}, M.}
\newblock \bibinfo{title}{{Predicted diversity in water content of terrestrial exoplanets orbiting M dwarfs}}.
\newblock \emph{\bibinfo{journal}{Nature Astronomy}} \textbf{\bibinfo{volume}{6}}, \bibinfo{pages}{1296--1307} (\bibinfo{year}{2022}).
\newblock \eprint{2209.14563}.

\bibitem{Luger2015}
\bibinfo{author}{{Luger}, R.} \& \bibinfo{author}{{Barnes}, R.}
\newblock \bibinfo{title}{{Extreme Water Loss and Abiotic O$_2$ Buildup on Planets Throughout the Habitable Zones of M Dwarfs}}.
\newblock \emph{\bibinfo{journal}{Astrobiology}} \textbf{\bibinfo{volume}{15}}, \bibinfo{pages}{119--143} (\bibinfo{year}{2015}).
\newblock \eprint{1411.7412}.

\bibitem{Dong2017}
\bibinfo{author}{{Dong}, C.} \emph{et~al.}
\newblock \bibinfo{title}{{The Dehydration of Water Worlds via Atmospheric Losses}}.
\newblock \emph{\bibinfo{journal}{\apjl}} \textbf{\bibinfo{volume}{847}}, \bibinfo{pages}{L4} (\bibinfo{year}{2017}).
\newblock \eprint{1709.01219}.

\bibitem{lincowski2018}
\bibinfo{author}{Lincowski, A.~P.} \emph{et~al.}
\newblock \bibinfo{title}{Evolved climates and observational discriminants for the trappist-1 planetary system}.
\newblock \emph{\bibinfo{journal}{The Astrophysical Journal}} \textbf{\bibinfo{volume}{867}}, \bibinfo{pages}{76} (\bibinfo{year}{2018}).

\bibitem{Dong2018}
\bibinfo{author}{{Dong}, C.} \emph{et~al.}
\newblock \bibinfo{title}{{Atmospheric escape from the TRAPPIST-1 planets and implications for habitability}}.
\newblock \emph{\bibinfo{journal}{Proceedings of the National Academy of Science}} \textbf{\bibinfo{volume}{115}}, \bibinfo{pages}{260--265} (\bibinfo{year}{2018}).
\newblock \eprint{1705.05535}.

\bibitem{Kral2018}
\bibinfo{author}{{Kral}, Q.} \emph{et~al.}
\newblock \bibinfo{title}{{Cometary impactors on the TRAPPIST-1 planets can destroy all planetary atmospheres and rebuild secondary atmospheres on planets f, g, and h}}.
\newblock \emph{\bibinfo{journal}{\mnras}} \textbf{\bibinfo{volume}{479}}, \bibinfo{pages}{2649--2672} (\bibinfo{year}{2018}).
\newblock \eprint{1802.05034}.

\bibitem{Hu2023}
\bibinfo{author}{{Hu}, R.}, \bibinfo{author}{{Gaillard}, F.} \& \bibinfo{author}{{Kite}, E.~S.}
\newblock \bibinfo{title}{{Narrow Loophole for H$_{2}$-Dominated Atmospheres on Habitable Rocky Planets around M Dwarfs}}.
\newblock \emph{\bibinfo{journal}{\apjl}} \textbf{\bibinfo{volume}{948}}, \bibinfo{pages}{L20} (\bibinfo{year}{2023}).
\newblock \eprint{2304.13659}.

\bibitem{Kostov2019}
\bibinfo{author}{{Kostov}, V.~B.} \emph{et~al.}
\newblock \bibinfo{title}{{The L 98-59 System: Three Transiting, Terrestrial-size Planets Orbiting a Nearby M Dwarf}}.
\newblock \emph{\bibinfo{journal}{\aj}} \textbf{\bibinfo{volume}{158}}, \bibinfo{pages}{32} (\bibinfo{year}{2019}).
\newblock \eprint{1903.08017}.

\bibitem{Dittmann2017}
\bibinfo{author}{{Dittmann}, J.~A.} \emph{et~al.}
\newblock \bibinfo{title}{{A temperate rocky super-Earth transiting a nearby cool star}}.
\newblock \emph{\bibinfo{journal}{\nat}} \textbf{\bibinfo{volume}{544}}, \bibinfo{pages}{333--336} (\bibinfo{year}{2017}).
\newblock \eprint{1704.05556}.

\bibitem{Vanderspek2019}
\bibinfo{author}{{Vanderspek}, R.} \emph{et~al.}
\newblock \bibinfo{title}{{TESS Discovery of an Ultra-short-period Planet around the Nearby M Dwarf LHS 3844}}.
\newblock \emph{\bibinfo{journal}{\apjl}} \textbf{\bibinfo{volume}{871}}, \bibinfo{pages}{L24} (\bibinfo{year}{2019}).
\newblock \eprint{1809.07242}.

\bibitem{Crossfield2019}
\bibinfo{author}{{Crossfield}, I. J.~M.} \emph{et~al.}
\newblock \bibinfo{title}{{A Super-Earth and Sub-Neptune Transiting the Late-type M Dwarf LP 791-18}}.
\newblock \emph{\bibinfo{journal}{\apjl}} \textbf{\bibinfo{volume}{883}}, \bibinfo{pages}{L16} (\bibinfo{year}{2019}).
\newblock \eprint{1906.09267}.

\bibitem{Peterson2023}
\bibinfo{author}{{Peterson}, M.~S.} \emph{et~al.}
\newblock \bibinfo{title}{{A temperate Earth-sized planet with tidal heating transiting an M6 star}}.
\newblock \emph{\bibinfo{journal}{\nat}} \textbf{\bibinfo{volume}{617}}, \bibinfo{pages}{701--705} (\bibinfo{year}{2023}).

\bibitem{Delrez2022}
\bibinfo{author}{{Delrez}, L.} \emph{et~al.}
\newblock \bibinfo{title}{{Two temperate super-Earths transiting a nearby late-type M dwarf}}.
\newblock \emph{\bibinfo{journal}{\aap}} \textbf{\bibinfo{volume}{667}}, \bibinfo{pages}{A59} (\bibinfo{year}{2022}).
\newblock \eprint{2209.02831}.

\bibitem{Ment2021}
\bibinfo{author}{{Ment}, K.} \emph{et~al.}
\newblock \bibinfo{title}{{TOI 540 b: A Planet Smaller than Earth Orbiting a Nearby Rapidly Rotating Low-mass Star}}.
\newblock \emph{\bibinfo{journal}{\aj}} \textbf{\bibinfo{volume}{161}}, \bibinfo{pages}{23} (\bibinfo{year}{2021}).
\newblock \eprint{2009.13623}.

\bibitem{Gillon2016}
\bibinfo{author}{{Gillon}, M.} \emph{et~al.}
\newblock \bibinfo{title}{{Temperate Earth-sized planets transiting a nearby ultracool dwarf star}}.
\newblock \emph{\bibinfo{journal}{Nature}} \textbf{\bibinfo{volume}{533}}, \bibinfo{pages}{221--224} (\bibinfo{year}{2016}).
\newblock \eprint{1605.07211}.

\bibitem{Gillon2017}
\bibinfo{author}{Gillon, M.} \emph{et~al.}
\newblock \bibinfo{title}{Seven temperate terrestrial planets around the nearby ultracool dwarf star {TRAPPIST}-1}.
\newblock \emph{\bibinfo{journal}{Nature}} \textbf{\bibinfo{volume}{542}}, \bibinfo{pages}{456--460} (\bibinfo{year}{2017}).
\newblock \urlprefix\url{https://doi.org/10.1038/nature21360}.

\bibitem{Ormel2017}
\bibinfo{author}{{Ormel}, C.~W.}, \bibinfo{author}{{Liu}, B.} \& \bibinfo{author}{{Schoonenberg}, D.}
\newblock \bibinfo{title}{{Formation of TRAPPIST-1 and other compact systems}}.
\newblock \emph{\bibinfo{journal}{\aap}} \textbf{\bibinfo{volume}{604}}, \bibinfo{pages}{A1} (\bibinfo{year}{2017}).
\newblock \eprint{1703.06924}.

\bibitem{Weiss2018}
\bibinfo{author}{{Weiss}, L.~M.} \emph{et~al.}
\newblock \bibinfo{title}{{The California-Kepler Survey. V. Peas in a Pod: Planets in a Kepler Multi-planet System Are Similar in Size and Regularly Spaced}}.
\newblock \emph{\bibinfo{journal}{\aj}} \textbf{\bibinfo{volume}{155}}, \bibinfo{pages}{48} (\bibinfo{year}{2018}).
\newblock \eprint{1706.06204}.

\bibitem{Sandford2021}
\bibinfo{author}{{Sandford}, E.}, \bibinfo{author}{{Kipping}, D.} \& \bibinfo{author}{{Collins}, M.}
\newblock \bibinfo{title}{{On planetary systems as ordered sequences}}.
\newblock \emph{\bibinfo{journal}{\mnras}} \textbf{\bibinfo{volume}{505}}, \bibinfo{pages}{2224--2246} (\bibinfo{year}{2021}).
\newblock \eprint{2105.09966}.

\bibitem{Mishra2021}
\bibinfo{author}{{Mishra}, L.} \emph{et~al.}
\newblock \bibinfo{title}{{The New Generation Planetary Population Synthesis (NGPPS) VI. Introducing KOBE: Kepler Observes Bern Exoplanets. Theoretical perspectives on the architecture of planetary systems: Peas in a pod}}.
\newblock \emph{\bibinfo{journal}{\aap}} \textbf{\bibinfo{volume}{656}}, \bibinfo{pages}{A74} (\bibinfo{year}{2021}).
\newblock \eprint{2105.12745}.

\bibitem{Millholland2021}
\bibinfo{author}{{Millholland}, S.~C.} \& \bibinfo{author}{{Winn}, J.~N.}
\newblock \bibinfo{title}{{Split Peas in a Pod: Intra-system Uniformity of Super-Earths and Sub-Neptunes}}.
\newblock \emph{\bibinfo{journal}{\apjl}} \textbf{\bibinfo{volume}{920}}, \bibinfo{pages}{L34} (\bibinfo{year}{2021}).
\newblock \eprint{2110.01466}.

\bibitem{Goyal2022}
\bibinfo{author}{{Goyal}, A.~V.} \& \bibinfo{author}{{Wang}, S.}
\newblock \bibinfo{title}{{Generalized Peas in a Pod: Extending Intra-system Mass Uniformity to Non-TTV Systems via the Gini Index}}.
\newblock \emph{\bibinfo{journal}{\apj}} \textbf{\bibinfo{volume}{933}}, \bibinfo{pages}{162} (\bibinfo{year}{2022}).
\newblock \eprint{2206.00053}.

\bibitem{Delrez2018}
\bibinfo{author}{{Delrez}, L.} \emph{et~al.}
\newblock \bibinfo{title}{{SPECULOOS: a network of robotic telescopes to hunt for terrestrial planets around the nearest ultracool dwarfs}}.
\newblock In \bibinfo{editor}{{Marshall}, H.~K.} \& \bibinfo{editor}{{Spyromilio}, J.} (eds.) \emph{\bibinfo{booktitle}{Ground-based and Airborne Telescopes VII}}, vol. \bibinfo{volume}{10700} of \emph{\bibinfo{series}{Society of Photo-Optical Instrumentation Engineers (SPIE) Conference Series}}, \bibinfo{pages}{107001I} (\bibinfo{year}{2018}).
\newblock \eprint{1806.11205}.

\bibitem{Burdanov2018}
\bibinfo{author}{{Burdanov}, A.}, \bibinfo{author}{{Delrez}, L.}, \bibinfo{author}{{Gillon}, M.} \& \bibinfo{author}{{Jehin}, E.}
\newblock \bibinfo{title}{{SPECULOOS Exoplanet Search and Its Prototype on TRAPPIST}}.
\newblock In \bibinfo{editor}{{Deeg}, H.~J.} \& \bibinfo{editor}{{Belmonte}, J.~A.} (eds.) \emph{\bibinfo{booktitle}{Handbook of Exoplanets}}, \bibinfo{pages}{130} (\bibinfo{year}{2018}).

\bibitem{Gibbs2020}
\bibinfo{author}{{Gibbs}, A.} \emph{et~al.}
\newblock \bibinfo{title}{{EDEN: Sensitivity Analysis and Transiting Planet Detection Limits for Nearby Late Red Dwarfs}}.
\newblock \emph{\bibinfo{journal}{\aj}} \textbf{\bibinfo{volume}{159}}, \bibinfo{pages}{169} (\bibinfo{year}{2020}).
\newblock \eprint{2002.10017}.

\bibitem{Tamburo2022}
\bibinfo{author}{{Tamburo}, P.} \emph{et~al.}
\newblock \bibinfo{title}{{The Perkins INfrared Exosatellite Survey (PINES) I. Survey Overview, Reduction Pipeline, and Early Results}}.
\newblock \emph{\bibinfo{journal}{\aj}} \textbf{\bibinfo{volume}{163}}, \bibinfo{pages}{253} (\bibinfo{year}{2022}).
\newblock \eprint{2201.01794}.

\bibitem{Jehin2011}
\bibinfo{author}{{Jehin}, E.} \emph{et~al.}
\newblock \bibinfo{title}{{TRAPPIST: TRAnsiting Planets and PlanetesImals Small Telescope}}.
\newblock \emph{\bibinfo{journal}{The Messenger}} \textbf{\bibinfo{volume}{145}}, \bibinfo{pages}{2--6} (\bibinfo{year}{2011}).

\bibitem{Gillon2020}
\bibinfo{author}{{Gillon}, M.} \emph{et~al.}
\newblock \bibinfo{title}{{The TRAPPIST-1 JWST Community Initiative}}.
\newblock In \emph{\bibinfo{booktitle}{Bulletin of the American Astronomical Society}}, vol.~\bibinfo{volume}{52}, \bibinfo{pages}{0208} (\bibinfo{year}{2020}).

\bibitem{Ducrot2020}
\bibinfo{author}{{Ducrot}, E.} \emph{et~al.}
\newblock \bibinfo{title}{{TRAPPIST-1: Global results of the Spitzer Exploration Science Program Red Worlds}}.
\newblock \emph{\bibinfo{journal}{\aap}} \textbf{\bibinfo{volume}{640}}, \bibinfo{pages}{A112} (\bibinfo{year}{2020}).
\newblock \eprint{2006.13826}.

\bibitem{Wakeford2019}
\bibinfo{author}{{Wakeford}, H.~R.} \emph{et~al.}
\newblock \bibinfo{title}{{Disentangling the Planet from the Star in Late-Type M Dwarfs: A Case Study of TRAPPIST-1g}}.
\newblock \emph{\bibinfo{journal}{\aj}} \textbf{\bibinfo{volume}{157}}, \bibinfo{pages}{11} (\bibinfo{year}{2019}).
\newblock \eprint{1811.04877}.

\bibitem{Turbet2020}
\bibinfo{author}{{Turbet}, M.} \emph{et~al.}
\newblock \bibinfo{title}{{A Review of Possible Planetary Atmospheres in the TRAPPIST-1 System}}.
\newblock \emph{\bibinfo{journal}{\ssr}} \textbf{\bibinfo{volume}{216}}, \bibinfo{pages}{100} (\bibinfo{year}{2020}).
\newblock \eprint{2007.03334}.

\bibitem{Agol2021}
\bibinfo{author}{{Agol}, E.} \emph{et~al.}
\newblock \bibinfo{title}{{Refining the Transit-timing and Photometric Analysis of TRAPPIST-1: Masses, Radii, Densities, Dynamics, and Ephemerides}}.
\newblock \emph{\bibinfo{journal}{\psj}} \textbf{\bibinfo{volume}{2}}, \bibinfo{pages}{1} (\bibinfo{year}{2021}).
\newblock \eprint{2010.01074}.

\bibitem{Garcia2022}
\bibinfo{author}{{Garcia}, L.~J.} \emph{et~al.}
\newblock \bibinfo{title}{{HST/WFC3 transmission spectroscopy of the cold rocky planet TRAPPIST-1h}}.
\newblock \emph{\bibinfo{journal}{\aap}} \textbf{\bibinfo{volume}{665}}, \bibinfo{pages}{A19} (\bibinfo{year}{2022}).
\newblock \eprint{2203.13698}.

\bibitem{Gressier2022}
\bibinfo{author}{{Gressier}, A.} \emph{et~al.}
\newblock \bibinfo{title}{{Near-infrared transmission spectrum of TRAPPIST-1 h using Hubble WFC3 G141 observations}}.
\newblock \emph{\bibinfo{journal}{\aap}} \textbf{\bibinfo{volume}{658}}, \bibinfo{pages}{A133} (\bibinfo{year}{2022}).
\newblock \eprint{2112.05510}.

\bibitem{Dorn2021}
\bibinfo{author}{{Dorn}, C.} \& \bibinfo{author}{{Lichtenberg}, T.}
\newblock \bibinfo{title}{{Hidden Water in Magma Ocean Exoplanets}}.
\newblock \emph{\bibinfo{journal}{\apjl}} \textbf{\bibinfo{volume}{922}}, \bibinfo{pages}{L4} (\bibinfo{year}{2021}).
\newblock \eprint{2110.15069}.

\bibitem{Benneke2023}
\bibinfo{author}{{Benneke}, B. e.~a.}
\newblock \bibinfo{title}{{JWST NIRSpec Reconnaissance Transmission Spectroscopy of the Habitable-zone Exo-Earth TRAPPIST-1~g}}.
\newblock \emph{\bibinfo{journal}{in prep.}}  (\bibinfo{year}{2023}).

\bibitem{Moran2023}
\bibinfo{author}{{Moran}, S.~E.} \emph{et~al.}
\newblock \bibinfo{title}{{High Tide or Riptide on the Cosmic Shoreline? A Water-Rich Atmosphere or Stellar Contamination for the Warm Super-Earth GJ\raisebox{-0.5ex}\textasciitilde486b from JWST Observations}}.
\newblock \emph{\bibinfo{journal}{arXiv e-prints}} \bibinfo{pages}{arXiv:2305.00868} (\bibinfo{year}{2023}).
\newblock \eprint{2305.00868}.

\bibitem{rackham2018}
\bibinfo{author}{{Rackham}, B.~V.}, \bibinfo{author}{{Apai}, D.} \& \bibinfo{author}{{Giampapa}, M.~S.}
\newblock \bibinfo{title}{{The Transit Light Source Effect: False Spectral Features and Incorrect Densities for M-dwarf Transiting Planets}}.
\newblock \emph{\bibinfo{journal}{\apj}} \textbf{\bibinfo{volume}{853}}, \bibinfo{pages}{122} (\bibinfo{year}{2018}).
\newblock \eprint{1711.05691}.

\bibitem{Rackham2019A}
\bibinfo{author}{{Rackham}, B.~V.}, \bibinfo{author}{{Apai}, D.} \& \bibinfo{author}{{Giampapa}, M.~S.}
\newblock \bibinfo{title}{{The Transit Light Source Effect. II. The Impact of Stellar Heterogeneity on Transmission Spectra of Planets Orbiting Broadly Sun-like Stars}}.
\newblock \emph{\bibinfo{journal}{\aj}} \textbf{\bibinfo{volume}{157}}, \bibinfo{pages}{96} (\bibinfo{year}{2019}).
\newblock \eprint{1812.06184}.

\bibitem{Witzke2021}
\bibinfo{author}{{Witzke}, V.} \emph{et~al.}
\newblock \bibinfo{title}{{MPS-ATLAS: A fast all-in-one code for synthesising stellar spectra}}.
\newblock \emph{\bibinfo{journal}{\aap}} \textbf{\bibinfo{volume}{653}}, \bibinfo{pages}{A65} (\bibinfo{year}{2021}).
\newblock \eprint{2105.13611}.

\bibitem{Rustamkulov2022}
\bibinfo{author}{{Rustamkulov}, Z.} \emph{et~al.}
\newblock \bibinfo{title}{{Early Release Science of the exoplanet WASP-39b with JWST NIRSpec PRISM}}.
\newblock \emph{\bibinfo{journal}{arXiv e-prints}} \bibinfo{pages}{arXiv:2211.10487} (\bibinfo{year}{2022}).
\newblock \eprint{2211.10487}.

\bibitem{Zhang2018}
\bibinfo{author}{{Zhang}, Z.}, \bibinfo{author}{{Zhou}, Y.}, \bibinfo{author}{{Rackham}, B.~V.} \& \bibinfo{author}{{Apai}, D.}
\newblock \bibinfo{title}{{The Near-infrared Transmission Spectra of TRAPPIST-1 Planets b, c, d, e, f, and g and Stellar Contamination in Multi-epoch Transit Spectra}}.
\newblock \emph{\bibinfo{journal}{The Astronomical Journal}} \textbf{\bibinfo{volume}{156}}, \bibinfo{pages}{178} (\bibinfo{year}{2018}).
\newblock \eprint{1802.02086}.

\bibitem{Morley2017}
\bibinfo{author}{{Morley}, C.~V.}, \bibinfo{author}{{Kreidberg}, L.}, \bibinfo{author}{{Rustamkulov}, Z.}, \bibinfo{author}{{Robinson}, T.} \& \bibinfo{author}{{Fortney}, J.~J.}
\newblock \bibinfo{title}{{Observing the Atmospheres of Known Temperate Earth-sized Planets with JWST}}.
\newblock \emph{\bibinfo{journal}{\apj}} \textbf{\bibinfo{volume}{850}}, \bibinfo{pages}{121} (\bibinfo{year}{2017}).
\newblock \eprint{1708.04239}.

\bibitem{Lustig-Yaeger2019}
\bibinfo{author}{{Lustig-Yaeger}, J.}, \bibinfo{author}{{Meadows}, V.~S.} \& \bibinfo{author}{{Lincowski}, A.~P.}
\newblock \bibinfo{title}{{The Detectability and Characterization of the TRAPPIST-1 Exoplanet Atmospheres with JWST}}.
\newblock \emph{\bibinfo{journal}{The Astronomical Journal}} \textbf{\bibinfo{volume}{158}}, \bibinfo{pages}{27} (\bibinfo{year}{2019}).
\newblock \eprint{1905.07070}.

\bibitem{Krissansen2018}
\bibinfo{author}{{Krissansen-Totton}, J.}, \bibinfo{author}{{Garland}, R.}, \bibinfo{author}{{Irwin}, P.} \& \bibinfo{author}{{Catling}, D.~C.}
\newblock \bibinfo{title}{{Detectability of Biosignatures in Anoxic Atmospheres with the James Webb Space Telescope: A TRAPPIST-1e Case Study}}.
\newblock \emph{\bibinfo{journal}{\aj}} \textbf{\bibinfo{volume}{156}}, \bibinfo{pages}{114} (\bibinfo{year}{2018}).
\newblock \eprint{1808.08377}.

\bibitem{Fauchez2019}
\bibinfo{author}{{Fauchez}, T.~J.} \emph{et~al.}
\newblock \bibinfo{title}{{Impact of Clouds and Hazes on the Simulated JWST Transmission Spectra of Habitable Zone Planets in the TRAPPIST-1 System}}.
\newblock \emph{\bibinfo{journal}{\apj}} \textbf{\bibinfo{volume}{887}}, \bibinfo{pages}{194} (\bibinfo{year}{2019}).
\newblock \eprint{1911.08596}.

\bibitem{Wunderlich2019}
\bibinfo{author}{{Wunderlich}, F.} \emph{et~al.}
\newblock \bibinfo{title}{{Detectability of atmospheric features of Earth-like planets in the habitable zone around M dwarfs}}.
\newblock \emph{\bibinfo{journal}{\aap}} \textbf{\bibinfo{volume}{624}}, \bibinfo{pages}{A49} (\bibinfo{year}{2019}).
\newblock \eprint{1905.02560}.

\bibitem{Gialluca2021}
\bibinfo{author}{{Gialluca}, M.~T.}, \bibinfo{author}{{Robinson}, T.~D.}, \bibinfo{author}{{Rugheimer}, S.} \& \bibinfo{author}{{Wunderlich}, F.}
\newblock \bibinfo{title}{{Characterizing Atmospheres of Transiting Earth-like Exoplanets Orbiting M Dwarfs with James Webb Space Telescope}}.
\newblock \emph{\bibinfo{journal}{\pasp}} \textbf{\bibinfo{volume}{133}}, \bibinfo{pages}{054401} (\bibinfo{year}{2021}).
\newblock \eprint{2101.04139}.

\bibitem{Faria2022}
\bibinfo{author}{{Faria}, J.~P.} \emph{et~al.}
\newblock \bibinfo{title}{{A candidate short-period sub-Earth orbiting Proxima Centauri}}.
\newblock \emph{\bibinfo{journal}{\aap}} \textbf{\bibinfo{volume}{658}}, \bibinfo{pages}{A115} (\bibinfo{year}{2022}).
\newblock \eprint{2202.05188}.

\bibitem{Segura2010}
\bibinfo{author}{{Segura}, A.}, \bibinfo{author}{{Walkowicz}, L.~M.}, \bibinfo{author}{{Meadows}, V.}, \bibinfo{author}{{Kasting}, J.} \& \bibinfo{author}{{Hawley}, S.}
\newblock \bibinfo{title}{{The Effect of a Strong Stellar Flare on the Atmospheric Chemistry of an Earth-like Planet Orbiting an M Dwarf}}.
\newblock \emph{\bibinfo{journal}{Astrobiology}} \textbf{\bibinfo{volume}{10}}, \bibinfo{pages}{751--771} (\bibinfo{year}{2010}).
\newblock \eprint{1006.0022}.

\bibitem{Grayver2022}
\bibinfo{author}{{Grayver}, A.}, \bibinfo{author}{{Bower}, D.~J.}, \bibinfo{author}{{Saur}, J.}, \bibinfo{author}{{Dorn}, C.} \& \bibinfo{author}{{Morris}, B.~M.}
\newblock \bibinfo{title}{{Interior Heating of Rocky Exoplanets from Stellar Flares with Application to TRAPPIST-1}}.
\newblock \emph{\bibinfo{journal}{\apjl}} \textbf{\bibinfo{volume}{941}}, \bibinfo{pages}{L7} (\bibinfo{year}{2022}).
\newblock \eprint{2211.06140}.

\bibitem{Ilin2021}
\bibinfo{author}{{Ilin}, E.} \emph{et~al.}
\newblock \bibinfo{title}{{Giant white-light flares on fully convective stars occur at high latitudes}}.
\newblock \emph{\bibinfo{journal}{\mnras}} \textbf{\bibinfo{volume}{507}}, \bibinfo{pages}{1723--1745} (\bibinfo{year}{2021}).
\newblock \eprint{2108.01917}.

\bibitem{Luger2021a}
\bibinfo{author}{{Luger}, R.}, \bibinfo{author}{{Foreman-Mackey}, D.}, \bibinfo{author}{{Hedges}, C.} \& \bibinfo{author}{{Hogg}, D.~W.}
\newblock \bibinfo{title}{{Mapping Stellar Surfaces. I. Degeneracies in the Rotational Light-curve Problem}}.
\newblock \emph{\bibinfo{journal}{\aj}} \textbf{\bibinfo{volume}{162}}, \bibinfo{pages}{123} (\bibinfo{year}{2021}).
\newblock \eprint{2102.00007}.

\bibitem{Luger2021b}
\bibinfo{author}{{Luger}, R.}, \bibinfo{author}{{Foreman-Mackey}, D.} \& \bibinfo{author}{{Hedges}, C.}
\newblock \bibinfo{title}{{Mapping Stellar Surfaces. II. An Interpretable Gaussian Process Model for Light Curves}}.
\newblock \emph{\bibinfo{journal}{\aj}} \textbf{\bibinfo{volume}{162}}, \bibinfo{pages}{124} (\bibinfo{year}{2021}).
\newblock \eprint{2102.01697}.

\bibitem{Luger2022}
\bibinfo{author}{{Luger}, R.} \emph{et~al.}
\newblock \bibinfo{title}{{Mapping stellar surfaces III: An Efficient, Scalable, and Open-Source Doppler Imaging Model}}.
\newblock \emph{\bibinfo{journal}{arXiv e-prints}} \bibinfo{pages}{arXiv:2110.06271} (\bibinfo{year}{2021}).
\newblock \eprint{2110.06271}.

\bibitem{Bolmont2017}
\bibinfo{author}{{Bolmont}, E.} \emph{et~al.}
\newblock \bibinfo{title}{{Water loss from terrestrial planets orbiting ultracool dwarfs: implications for the planets of TRAPPIST-1}}.
\newblock \emph{\bibinfo{journal}{Monthly Notices of the Royal Astronomical Society}} \textbf{\bibinfo{volume}{464}}, \bibinfo{pages}{3728--3741} (\bibinfo{year}{2017}).
\newblock \eprint{1605.00616}.

\bibitem{Bourrier2017b}
\bibinfo{author}{{Bourrier}, V.} \emph{et~al.}
\newblock \bibinfo{title}{{Temporal Evolution of the High-energy Irradiation and Water Content of TRAPPIST-1 Exoplanets}}.
\newblock \emph{\bibinfo{journal}{The Astronomical Journal}} \textbf{\bibinfo{volume}{154}}, \bibinfo{pages}{121} (\bibinfo{year}{2017}).
\newblock \eprint{1708.09484}.

\bibitem{Rimmer2018}
\bibinfo{author}{{Rimmer}, P.~B.} \emph{et~al.}
\newblock \bibinfo{title}{{The origin of RNA precursors on exoplanets}}.
\newblock \emph{\bibinfo{journal}{Science Advances}} \textbf{\bibinfo{volume}{4}}, \bibinfo{pages}{eaar3302} (\bibinfo{year}{2018}).
\newblock \eprint{1808.02718}.

\bibitem{Rackham2023}
\bibinfo{author}{{Rackham}, B.~V.} \& \bibinfo{author}{{de Wit}, J.}
\newblock \bibinfo{title}{{Towards robust corrections for stellar contamination in JWST exoplanet transmission spectra}}.
\newblock \emph{\bibinfo{journal}{arXiv e-prints}} \bibinfo{pages}{arXiv:2303.15418} (\bibinfo{year}{2023}).
\newblock \eprint{2303.15418}.

\bibitem{SAG21_RASTI}
\bibinfo{author}{{Rackham}, B.~V.} \& \bibinfo{author}{{de Wit}, J.}
\newblock \bibinfo{title}{{Towards robust corrections for stellar contamination in JWST exoplanet transmission spectra}}.
\newblock \emph{\bibinfo{journal}{arXiv e-prints}} \bibinfo{pages}{arXiv:2303.15418} (\bibinfo{year}{2023}).
\newblock \eprint{2303.15418}.

\bibitem{Vogler2005}
\bibinfo{author}{{V{\"o}gler}, A.} \emph{et~al.}
\newblock \bibinfo{title}{{Simulations of magneto-convection in the solar photosphere. Equations, methods, and results of the MURaM code}}.
\newblock \emph{\bibinfo{journal}{\aap}} \textbf{\bibinfo{volume}{429}}, \bibinfo{pages}{335--351} (\bibinfo{year}{2005}).

\bibitem{Berardo2023b}
\bibinfo{author}{{Berardo}, D.}, \bibinfo{author}{{de Wit}, J.} \& \bibinfo{author}{{Rackham}, B.~V.}
\newblock \bibinfo{title}{{Empirically Constraining the Spectra of a Stars Heterogeneities From Its Rotation Lightcurve}}.
\newblock \emph{\bibinfo{journal}{arXiv e-prints}} \bibinfo{pages}{arXiv:2307.04785} (\bibinfo{year}{2023}).
\newblock \eprint{2307.04785}.

\bibitem{Morris2018}
\bibinfo{author}{Morris, B.~M.} \emph{et~al.}
\newblock \bibinfo{title}{Non-detection of contamination by stellar activity in the spitzer transit light curves of {TRAPPIST}-1}.
\newblock \emph{\bibinfo{journal}{The Astrophysical Journal}} \textbf{\bibinfo{volume}{863}}, \bibinfo{pages}{L32} (\bibinfo{year}{2018}).
\newblock \urlprefix\url{https://doi.org/10.3847/2041-8213/aad8aa}.

\bibitem{Krissansen2023}
\bibinfo{author}{{Krissansen-Totton}, J.}
\newblock \bibinfo{title}{{Implications of atmospheric non-detections for Trappist-1 inner planets on atmospheric retention prospects for outer planets}}.
\newblock \emph{\bibinfo{journal}{arXiv e-prints}} \bibinfo{pages}{arXiv:2306.05397} (\bibinfo{year}{2023}).
\newblock \eprint{2306.05397}.

\bibitem{Mallonn2018}
\bibinfo{author}{{Mallonn}, M.} \emph{et~al.}
\newblock \bibinfo{title}{{GJ 1214: Rotation period, starspots, and uncertainty on the optical slope of the transmission spectrum}}.
\newblock \emph{\bibinfo{journal}{\aap}} \textbf{\bibinfo{volume}{614}}, \bibinfo{pages}{A35} (\bibinfo{year}{2018}).
\newblock \eprint{1803.05677}.

\bibitem{Rosich2020}
\bibinfo{author}{{Rosich}, A.} \emph{et~al.}
\newblock \bibinfo{title}{{Correcting for chromatic stellar activity effects in transits with multiband photometric monitoring: application to WASP-52}}.
\newblock \emph{\bibinfo{journal}{\aap}} \textbf{\bibinfo{volume}{641}}, \bibinfo{pages}{A82} (\bibinfo{year}{2020}).
\newblock \eprint{2007.00573}.

\bibitem{Perger2023}
\bibinfo{author}{{Perger}, M.} \emph{et~al.}
\newblock \bibinfo{title}{{A machine learning approach for correcting radial velocities using physical observables}}.
\newblock \emph{\bibinfo{journal}{\aap}} \textbf{\bibinfo{volume}{672}}, \bibinfo{pages}{A118} (\bibinfo{year}{2023}).
\newblock \eprint{2301.12872}.

\bibitem{Reyle2021}
\bibinfo{author}{{Reyl{\'e}}, C.} \emph{et~al.}
\newblock \bibinfo{title}{{The 10 parsec sample in the Gaia era}}.
\newblock \emph{\bibinfo{journal}{\aap}} \textbf{\bibinfo{volume}{650}}, \bibinfo{pages}{A201} (\bibinfo{year}{2021}).
\newblock \eprint{2104.14972}.

\bibitem{Kempton2018}
\bibinfo{author}{{Kempton}, E. M.~R.} \emph{et~al.}
\newblock \bibinfo{title}{{A Framework for Prioritizing the TESS Planetary Candidates Most Amenable to Atmospheric Characterization}}.
\newblock \emph{\bibinfo{journal}{\pasp}} \textbf{\bibinfo{volume}{130}}, \bibinfo{pages}{114401} (\bibinfo{year}{2018}).
\newblock \eprint{1805.03671}.

\bibitem{Morris2020}
\bibinfo{author}{{Morris}, B.}
\newblock \bibinfo{title}{{fleck: Fast approximate light curves for starspot rotational modulation}}.
\newblock \emph{\bibinfo{journal}{The Journal of Open Source Software}} \textbf{\bibinfo{volume}{5}}, \bibinfo{pages}{2103} (\bibinfo{year}{2020}).

\end{thebibliography}


\begin{thebibliography}{10}
\expandafter\ifx\csname url\endcsname\relax
  \def\url#1{\texttt{#1}}\fi
\expandafter\ifx\csname urlprefix\endcsname\relax\def\urlprefix{URL }\fi
\providecommand{\bibinfo}[2]{#2}
\providecommand{\eprint}[2][]{\url{#2}}

\bibitem{Dressing2015}
\bibinfo{author}{{Dressing}, C.~D.} \& \bibinfo{author}{{Charbonneau}, D.}
\newblock \bibinfo{title}{{The Occurrence of Potentially Habitable Planets Orbiting M Dwarfs Estimated from the Full Kepler Dataset and an Empirical Measurement of the Detection Sensitivity}}.
\newblock \emph{\bibinfo{journal}{\apj}} \textbf{\bibinfo{volume}{807}}, \bibinfo{pages}{45} (\bibinfo{year}{2015}).
\newblock \eprint{1501.01623}.

\bibitem{Gaidos2016}
\bibinfo{author}{{Gaidos}, E.}, \bibinfo{author}{{Mann}, A.~W.}, \bibinfo{author}{{Kraus}, A.~L.} \& \bibinfo{author}{{Ireland}, M.}
\newblock \bibinfo{title}{{They are small worlds after all: revised properties of Kepler M dwarf stars and their planets}}.
\newblock \emph{\bibinfo{journal}{\mnras}} \textbf{\bibinfo{volume}{457}}, \bibinfo{pages}{2877--2899} (\bibinfo{year}{2016}).
\newblock \eprint{1512.04437}.

\bibitem{Ment2023}
\bibinfo{author}{{Ment}, K.} \& \bibinfo{author}{{Charbonneau}, D.}
\newblock \bibinfo{title}{{The Occurrence Rate of Terrestrial Planets Orbiting Nearby Mid-to-late M Dwarfs from TESS Sectors 1-42}}.
\newblock \emph{\bibinfo{journal}{\aj}} \textbf{\bibinfo{volume}{165}}, \bibinfo{pages}{265} (\bibinfo{year}{2023}).
\newblock \eprint{2302.04242}.

\bibitem{Bochanski2010}
\bibinfo{author}{{Bochanski}, J.~J.} \emph{et~al.}
\newblock \bibinfo{title}{{The Luminosity and Mass Functions of Low-mass Stars in the Galactic Disk. II. The Field}}.
\newblock \emph{\bibinfo{journal}{\aj}} \textbf{\bibinfo{volume}{139}}, \bibinfo{pages}{2679--2699} (\bibinfo{year}{2010}).
\newblock \eprint{1004.4002}.

\bibitem{Triaud2023}
\bibinfo{author}{{Triaud}, A. H.~M.~J.} \emph{et~al.}
\newblock \bibinfo{title}{{Atmospheric carbon depletion as a tracer of water oceans and biomass on temperate terrestrial exoplanets}}.
\newblock \emph{\bibinfo{journal}{Nature Astronomy}}  (\bibinfo{year}{2023}).
\newblock \eprint{2310.14987}.

\bibitem{Kostov2019}
\bibinfo{author}{{Kostov}, V.~B.} \emph{et~al.}
\newblock \bibinfo{title}{{The L 98-59 System: Three Transiting, Terrestrial-size Planets Orbiting a Nearby M Dwarf}}.
\newblock \emph{\bibinfo{journal}{\aj}} \textbf{\bibinfo{volume}{158}}, \bibinfo{pages}{32} (\bibinfo{year}{2019}).
\newblock \eprint{1903.08017}.

\bibitem{Dittmann2017}
\bibinfo{author}{{Dittmann}, J.~A.} \emph{et~al.}
\newblock \bibinfo{title}{{A temperate rocky super-Earth transiting a nearby cool star}}.
\newblock \emph{\bibinfo{journal}{\nat}} \textbf{\bibinfo{volume}{544}}, \bibinfo{pages}{333--336} (\bibinfo{year}{2017}).
\newblock \eprint{1704.05556}.

\bibitem{Vanderspek2019}
\bibinfo{author}{{Vanderspek}, R.} \emph{et~al.}
\newblock \bibinfo{title}{{TESS Discovery of an Ultra-short-period Planet around the Nearby M Dwarf LHS 3844}}.
\newblock \emph{\bibinfo{journal}{\apjl}} \textbf{\bibinfo{volume}{871}}, \bibinfo{pages}{L24} (\bibinfo{year}{2019}).
\newblock \eprint{1809.07242}.

\bibitem{Crossfield2019}
\bibinfo{author}{{Crossfield}, I. J.~M.} \emph{et~al.}
\newblock \bibinfo{title}{{A Super-Earth and Sub-Neptune Transiting the Late-type M Dwarf LP 791-18}}.
\newblock \emph{\bibinfo{journal}{\apjl}} \textbf{\bibinfo{volume}{883}}, \bibinfo{pages}{L16} (\bibinfo{year}{2019}).
\newblock \eprint{1906.09267}.

\bibitem{Peterson2023}
\bibinfo{author}{{Peterson}, M.~S.} \emph{et~al.}
\newblock \bibinfo{title}{{A temperate Earth-sized planet with tidal heating transiting an M6 star}}.
\newblock \emph{\bibinfo{journal}{\nat}} \textbf{\bibinfo{volume}{617}}, \bibinfo{pages}{701--705} (\bibinfo{year}{2023}).

\bibitem{Delrez2022}
\bibinfo{author}{{Delrez}, L.} \emph{et~al.}
\newblock \bibinfo{title}{{Two temperate super-Earths transiting a nearby late-type M dwarf}}.
\newblock \emph{\bibinfo{journal}{\aap}} \textbf{\bibinfo{volume}{667}}, \bibinfo{pages}{A59} (\bibinfo{year}{2022}).
\newblock \eprint{2209.02831}.

\bibitem{Ment2021}
\bibinfo{author}{{Ment}, K.} \emph{et~al.}
\newblock \bibinfo{title}{{TOI 540 b: A Planet Smaller than Earth Orbiting a Nearby Rapidly Rotating Low-mass Star}}.
\newblock \emph{\bibinfo{journal}{\aj}} \textbf{\bibinfo{volume}{161}}, \bibinfo{pages}{23} (\bibinfo{year}{2021}).
\newblock \eprint{2009.13623}.

\bibitem{Gillon2016}
\bibinfo{author}{{Gillon}, M.} \emph{et~al.}
\newblock \bibinfo{title}{{Temperate Earth-sized planets transiting a nearby ultracool dwarf star}}.
\newblock \emph{\bibinfo{journal}{Nature}} \textbf{\bibinfo{volume}{533}}, \bibinfo{pages}{221--224} (\bibinfo{year}{2016}).
\newblock \eprint{1605.07211}.

\bibitem{Gillon2017}
\bibinfo{author}{Gillon, M.} \emph{et~al.}
\newblock \bibinfo{title}{Seven temperate terrestrial planets around the nearby ultracool dwarf star {TRAPPIST}-1}.
\newblock \emph{\bibinfo{journal}{Nature}} \textbf{\bibinfo{volume}{542}}, \bibinfo{pages}{456--460} (\bibinfo{year}{2017}).
\newblock \urlprefix\url{https://doi.org/10.1038/nature21360}.

\bibitem{Lichtenberg2022}
\bibinfo{author}{{Lichtenberg}, T.} \& \bibinfo{author}{{Clement}, M.~S.}
\newblock \bibinfo{title}{{Reduced Late Bombardment on Rocky Exoplanets around M Dwarfs}}.
\newblock \emph{\bibinfo{journal}{\apjl}} \textbf{\bibinfo{volume}{938}}, \bibinfo{pages}{L3} (\bibinfo{year}{2022}).
\newblock \eprint{2209.14037}.

\bibitem{Baraffe1998}
\bibinfo{author}{{Baraffe}, I.}, \bibinfo{author}{{Chabrier}, G.}, \bibinfo{author}{{Allard}, F.} \& \bibinfo{author}{{Hauschildt}, P.~H.}
\newblock \bibinfo{title}{{Evolutionary models for solar metallicity low-mass stars: mass-magnitude relationships and color-magnitude diagrams}}.
\newblock \emph{\bibinfo{journal}{\aap}} \textbf{\bibinfo{volume}{337}}, \bibinfo{pages}{403--412} (\bibinfo{year}{1998}).
\newblock \eprint{astro-ph/9805009}.

\bibitem{Baraffe2015}
\bibinfo{author}{{Baraffe}, I.}, \bibinfo{author}{{Homeier}, D.}, \bibinfo{author}{{Allard}, F.} \& \bibinfo{author}{{Chabrier}, G.}
\newblock \bibinfo{title}{{New evolutionary models for pre-main sequence and main sequence low-mass stars down to the hydrogen-burning limit}}.
\newblock \emph{\bibinfo{journal}{\aap}} \textbf{\bibinfo{volume}{577}}, \bibinfo{pages}{A42} (\bibinfo{year}{2015}).
\newblock \eprint{1503.04107}.

\bibitem{Kane2014}
\bibinfo{author}{{Kane}, S.~R.}, \bibinfo{author}{{Kopparapu}, R.~K.} \& \bibinfo{author}{{Domagal-Goldman}, S.~D.}
\newblock \bibinfo{title}{{On the Frequency of Potential Venus Analogs from Kepler Data}}.
\newblock \emph{\bibinfo{journal}{\apjl}} \textbf{\bibinfo{volume}{794}}, \bibinfo{pages}{L5} (\bibinfo{year}{2014}).
\newblock \eprint{1409.2886}.

\bibitem{Tian2015}
\bibinfo{author}{{Tian}, F.} \& \bibinfo{author}{{Ida}, S.}
\newblock \bibinfo{title}{{Water contents of Earth-mass planets around M dwarfs}}.
\newblock \emph{\bibinfo{journal}{Nature Geoscience}} \textbf{\bibinfo{volume}{8}}, \bibinfo{pages}{177--180} (\bibinfo{year}{2015}).

\bibitem{Kane2019}
\bibinfo{author}{{Kane}, S.~R.} \emph{et~al.}
\newblock \bibinfo{title}{{Venus as a Laboratory for Exoplanetary Science}}.
\newblock \emph{\bibinfo{journal}{Journal of Geophysical Research (Planets)}} \textbf{\bibinfo{volume}{124}}, \bibinfo{pages}{2015--2028} (\bibinfo{year}{2019}).
\newblock \eprint{1908.02783}.

\bibitem{Lichtenberg2019}
\bibinfo{author}{{Lichtenberg}, T.} \emph{et~al.}
\newblock \bibinfo{title}{{A water budget dichotomy of rocky protoplanets from $^{26}$Al-heating}}.
\newblock \emph{\bibinfo{journal}{Nature Astronomy}} \textbf{\bibinfo{volume}{3}}, \bibinfo{pages}{307--313} (\bibinfo{year}{2019}).
\newblock \eprint{1902.04026}.

\bibitem{Venturini2020}
\bibinfo{author}{{Venturini}, J.}, \bibinfo{author}{{Guilera}, O.~M.}, \bibinfo{author}{{Haldemann}, J.}, \bibinfo{author}{{Ronco}, M.~P.} \& \bibinfo{author}{{Mordasini}, C.}
\newblock \bibinfo{title}{{The nature of the radius valley. Hints from formation and evolution models}}.
\newblock \emph{\bibinfo{journal}{\aap}} \textbf{\bibinfo{volume}{643}}, \bibinfo{pages}{L1} (\bibinfo{year}{2020}).
\newblock \eprint{2008.05513}.

\bibitem{Way2020}
\bibinfo{author}{{Way}, M.~J.} \& \bibinfo{author}{{Del Genio}, A.~D.}
\newblock \bibinfo{title}{{Venusian Habitable Climate Scenarios: Modeling Venus Through Time and Applications to Slowly Rotating Venus-Like Exoplanets}}.
\newblock \emph{\bibinfo{journal}{Journal of Geophysical Research (Planets)}} \textbf{\bibinfo{volume}{125}}, \bibinfo{pages}{e06276} (\bibinfo{year}{2020}).
\newblock \eprint{2003.05704}.

\bibitem{Kimura2022}
\bibinfo{author}{{Kimura}, T.} \& \bibinfo{author}{{Ikoma}, M.}
\newblock \bibinfo{title}{{Predicted diversity in water content of terrestrial exoplanets orbiting M dwarfs}}.
\newblock \emph{\bibinfo{journal}{Nature Astronomy}} \textbf{\bibinfo{volume}{6}}, \bibinfo{pages}{1296--1307} (\bibinfo{year}{2022}).
\newblock \eprint{2209.14563}.

\bibitem{Luger2015}
\bibinfo{author}{{Luger}, R.} \& \bibinfo{author}{{Barnes}, R.}
\newblock \bibinfo{title}{{Extreme Water Loss and Abiotic O$_2$ Buildup on Planets Throughout the Habitable Zones of M Dwarfs}}.
\newblock \emph{\bibinfo{journal}{Astrobiology}} \textbf{\bibinfo{volume}{15}}, \bibinfo{pages}{119--143} (\bibinfo{year}{2015}).
\newblock \eprint{1411.7412}.

\bibitem{Dong2017}
\bibinfo{author}{{Dong}, C.} \emph{et~al.}
\newblock \bibinfo{title}{{The Dehydration of Water Worlds via Atmospheric Losses}}.
\newblock \emph{\bibinfo{journal}{\apjl}} \textbf{\bibinfo{volume}{847}}, \bibinfo{pages}{L4} (\bibinfo{year}{2017}).
\newblock \eprint{1709.01219}.

\bibitem{lincowski2018}
\bibinfo{author}{Lincowski, A.~P.} \emph{et~al.}
\newblock \bibinfo{title}{Evolved climates and observational discriminants for the trappist-1 planetary system}.
\newblock \emph{\bibinfo{journal}{The Astrophysical Journal}} \textbf{\bibinfo{volume}{867}}, \bibinfo{pages}{76} (\bibinfo{year}{2018}).

\bibitem{Dong2018}
\bibinfo{author}{{Dong}, C.} \emph{et~al.}
\newblock \bibinfo{title}{{Atmospheric escape from the TRAPPIST-1 planets and implications for habitability}}.
\newblock \emph{\bibinfo{journal}{Proceedings of the National Academy of Science}} \textbf{\bibinfo{volume}{115}}, \bibinfo{pages}{260--265} (\bibinfo{year}{2018}).
\newblock \eprint{1705.05535}.

\bibitem{Kral2018}
\bibinfo{author}{{Kral}, Q.} \emph{et~al.}
\newblock \bibinfo{title}{{Cometary impactors on the TRAPPIST-1 planets can destroy all planetary atmospheres and rebuild secondary atmospheres on planets f, g, and h}}.
\newblock \emph{\bibinfo{journal}{\mnras}} \textbf{\bibinfo{volume}{479}}, \bibinfo{pages}{2649--2672} (\bibinfo{year}{2018}).
\newblock \eprint{1802.05034}.

\bibitem{Hu2023}
\bibinfo{author}{{Hu}, R.}, \bibinfo{author}{{Gaillard}, F.} \& \bibinfo{author}{{Kite}, E.~S.}
\newblock \bibinfo{title}{{Narrow Loophole for H$_{2}$-Dominated Atmospheres on Habitable Rocky Planets around M Dwarfs}}.
\newblock \emph{\bibinfo{journal}{\apjl}} \textbf{\bibinfo{volume}{948}}, \bibinfo{pages}{L20} (\bibinfo{year}{2023}).
\newblock \eprint{2304.13659}.

\bibitem{Ormel2017}
\bibinfo{author}{{Ormel}, C.~W.}, \bibinfo{author}{{Liu}, B.} \& \bibinfo{author}{{Schoonenberg}, D.}
\newblock \bibinfo{title}{{Formation of TRAPPIST-1 and other compact systems}}.
\newblock \emph{\bibinfo{journal}{\aap}} \textbf{\bibinfo{volume}{604}}, \bibinfo{pages}{A1} (\bibinfo{year}{2017}).
\newblock \eprint{1703.06924}.

\bibitem{Weiss2018}
\bibinfo{author}{{Weiss}, L.~M.} \emph{et~al.}
\newblock \bibinfo{title}{{The California-Kepler Survey. V. Peas in a Pod: Planets in a Kepler Multi-planet System Are Similar in Size and Regularly Spaced}}.
\newblock \emph{\bibinfo{journal}{\aj}} \textbf{\bibinfo{volume}{155}}, \bibinfo{pages}{48} (\bibinfo{year}{2018}).
\newblock \eprint{1706.06204}.

\bibitem{Sandford2021}
\bibinfo{author}{{Sandford}, E.}, \bibinfo{author}{{Kipping}, D.} \& \bibinfo{author}{{Collins}, M.}
\newblock \bibinfo{title}{{On planetary systems as ordered sequences}}.
\newblock \emph{\bibinfo{journal}{\mnras}} \textbf{\bibinfo{volume}{505}}, \bibinfo{pages}{2224--2246} (\bibinfo{year}{2021}).
\newblock \eprint{2105.09966}.

\bibitem{Mishra2021}
\bibinfo{author}{{Mishra}, L.} \emph{et~al.}
\newblock \bibinfo{title}{{The New Generation Planetary Population Synthesis (NGPPS) VI. Introducing KOBE: Kepler Observes Bern Exoplanets. Theoretical perspectives on the architecture of planetary systems: Peas in a pod}}.
\newblock \emph{\bibinfo{journal}{\aap}} \textbf{\bibinfo{volume}{656}}, \bibinfo{pages}{A74} (\bibinfo{year}{2021}).
\newblock \eprint{2105.12745}.

\bibitem{Millholland2021}
\bibinfo{author}{{Millholland}, S.~C.} \& \bibinfo{author}{{Winn}, J.~N.}
\newblock \bibinfo{title}{{Split Peas in a Pod: Intra-system Uniformity of Super-Earths and Sub-Neptunes}}.
\newblock \emph{\bibinfo{journal}{\apjl}} \textbf{\bibinfo{volume}{920}}, \bibinfo{pages}{L34} (\bibinfo{year}{2021}).
\newblock \eprint{2110.01466}.

\bibitem{Goyal2022}
\bibinfo{author}{{Goyal}, A.~V.} \& \bibinfo{author}{{Wang}, S.}
\newblock \bibinfo{title}{{Generalized Peas in a Pod: Extending Intra-system Mass Uniformity to Non-TTV Systems via the Gini Index}}.
\newblock \emph{\bibinfo{journal}{\apj}} \textbf{\bibinfo{volume}{933}}, \bibinfo{pages}{162} (\bibinfo{year}{2022}).
\newblock \eprint{2206.00053}.

\bibitem{Delrez2018}
\bibinfo{author}{{Delrez}, L.} \emph{et~al.}
\newblock \bibinfo{title}{{SPECULOOS: a network of robotic telescopes to hunt for terrestrial planets around the nearest ultracool dwarfs}}.
\newblock In \bibinfo{editor}{{Marshall}, H.~K.} \& \bibinfo{editor}{{Spyromilio}, J.} (eds.) \emph{\bibinfo{booktitle}{Ground-based and Airborne Telescopes VII}}, vol. \bibinfo{volume}{10700} of \emph{\bibinfo{series}{Society of Photo-Optical Instrumentation Engineers (SPIE) Conference Series}}, \bibinfo{pages}{107001I} (\bibinfo{year}{2018}).
\newblock \eprint{1806.11205}.

\bibitem{Burdanov2018}
\bibinfo{author}{{Burdanov}, A.}, \bibinfo{author}{{Delrez}, L.}, \bibinfo{author}{{Gillon}, M.} \& \bibinfo{author}{{Jehin}, E.}
\newblock \bibinfo{title}{{SPECULOOS Exoplanet Search and Its Prototype on TRAPPIST}}.
\newblock In \bibinfo{editor}{{Deeg}, H.~J.} \& \bibinfo{editor}{{Belmonte}, J.~A.} (eds.) \emph{\bibinfo{booktitle}{Handbook of Exoplanets}}, \bibinfo{pages}{130} (\bibinfo{year}{2018}).

\bibitem{Gibbs2020}
\bibinfo{author}{{Gibbs}, A.} \emph{et~al.}
\newblock \bibinfo{title}{{EDEN: Sensitivity Analysis and Transiting Planet Detection Limits for Nearby Late Red Dwarfs}}.
\newblock \emph{\bibinfo{journal}{\aj}} \textbf{\bibinfo{volume}{159}}, \bibinfo{pages}{169} (\bibinfo{year}{2020}).
\newblock \eprint{2002.10017}.

\bibitem{Tamburo2022}
\bibinfo{author}{{Tamburo}, P.} \emph{et~al.}
\newblock \bibinfo{title}{{The Perkins INfrared Exosatellite Survey (PINES) I. Survey Overview, Reduction Pipeline, and Early Results}}.
\newblock \emph{\bibinfo{journal}{\aj}} \textbf{\bibinfo{volume}{163}}, \bibinfo{pages}{253} (\bibinfo{year}{2022}).
\newblock \eprint{2201.01794}.

\bibitem{Gillon2013a}
\bibinfo{author}{{Gillon}, M.}, \bibinfo{author}{{Jehin}, E.}, \bibinfo{author}{{Fumel}, A.}, \bibinfo{author}{{Magain}, P.} \& \bibinfo{author}{{Queloz}, D.}
\newblock \bibinfo{title}{{TRAPPIST-UCDTS: A prototype search for habitable planets transiting ultra-cool stars}}.
\newblock In \emph{\bibinfo{booktitle}{European Physical Journal Web of Conferences}}, vol.~\bibinfo{volume}{47} of \emph{\bibinfo{series}{European Physical Journal Web of Conferences}}, \bibinfo{pages}{03001} (\bibinfo{year}{2013}).

\bibitem{Ducrot2020}
\bibinfo{author}{{Ducrot}, E.} \emph{et~al.}
\newblock \bibinfo{title}{{TRAPPIST-1: Global results of the Spitzer Exploration Science Program Red Worlds}}.
\newblock \emph{\bibinfo{journal}{\aap}} \textbf{\bibinfo{volume}{640}}, \bibinfo{pages}{A112} (\bibinfo{year}{2020}).
\newblock \eprint{2006.13826}.

\bibitem{Agol2021}
\bibinfo{author}{{Agol}, E.} \emph{et~al.}
\newblock \bibinfo{title}{{Refining the Transit-timing and Photometric Analysis of TRAPPIST-1: Masses, Radii, Densities, Dynamics, and Ephemerides}}.
\newblock \emph{\bibinfo{journal}{\psj}} \textbf{\bibinfo{volume}{2}}, \bibinfo{pages}{1} (\bibinfo{year}{2021}).
\newblock \eprint{2010.01074}.

\bibitem{Dorn2021}
\bibinfo{author}{{Dorn}, C.} \& \bibinfo{author}{{Lichtenberg}, T.}
\newblock \bibinfo{title}{{Hidden Water in Magma Ocean Exoplanets}}.
\newblock \emph{\bibinfo{journal}{\apjl}} \textbf{\bibinfo{volume}{922}}, \bibinfo{pages}{L4} (\bibinfo{year}{2021}).
\newblock \eprint{2110.15069}.

\bibitem{dewit2016}
\bibinfo{author}{{de Wit}, J.} \emph{et~al.}
\newblock \bibinfo{title}{{A combined transmission spectrum of the Earth-sized exoplanets TRAPPIST-1 b and c}}.
\newblock \emph{\bibinfo{journal}{\nat}} \textbf{\bibinfo{volume}{537}}, \bibinfo{pages}{69--72} (\bibinfo{year}{2016}).
\newblock \eprint{1606.01103}.

\bibitem{deWit2018}
\bibinfo{author}{{de Wit}, J.} \emph{et~al.}
\newblock \bibinfo{title}{{Atmospheric reconnaissance of the habitable-zone Earth-sized planets orbiting TRAPPIST-1}}.
\newblock \emph{\bibinfo{journal}{Nature Astronomy}} \textbf{\bibinfo{volume}{2}}, \bibinfo{pages}{214--219} (\bibinfo{year}{2018}).
\newblock \eprint{1802.02250}.

\bibitem{Wakeford2019}
\bibinfo{author}{{Wakeford}, H.~R.} \emph{et~al.}
\newblock \bibinfo{title}{{Disentangling the Planet from the Star in Late-Type M Dwarfs: A Case Study of TRAPPIST-1g}}.
\newblock \emph{\bibinfo{journal}{\aj}} \textbf{\bibinfo{volume}{157}}, \bibinfo{pages}{11} (\bibinfo{year}{2019}).
\newblock \eprint{1811.04877}.

\bibitem{Turbet2020}
\bibinfo{author}{{Turbet}, M.} \emph{et~al.}
\newblock \bibinfo{title}{{A Review of Possible Planetary Atmospheres in the TRAPPIST-1 System}}.
\newblock \emph{\bibinfo{journal}{\ssr}} \textbf{\bibinfo{volume}{216}}, \bibinfo{pages}{100} (\bibinfo{year}{2020}).
\newblock \eprint{2007.03334}.

\bibitem{Garcia2022}
\bibinfo{author}{{Garcia}, L.~J.} \emph{et~al.}
\newblock \bibinfo{title}{{HST/WFC3 transmission spectroscopy of the cold rocky planet TRAPPIST-1h}}.
\newblock \emph{\bibinfo{journal}{\aap}} \textbf{\bibinfo{volume}{665}}, \bibinfo{pages}{A19} (\bibinfo{year}{2022}).
\newblock \eprint{2203.13698}.

\bibitem{Gressier2022}
\bibinfo{author}{{Gressier}, A.} \emph{et~al.}
\newblock \bibinfo{title}{{Near-infrared transmission spectrum of TRAPPIST-1 h using Hubble WFC3 G141 observations}}.
\newblock \emph{\bibinfo{journal}{\aap}} \textbf{\bibinfo{volume}{658}}, \bibinfo{pages}{A133} (\bibinfo{year}{2022}).
\newblock \eprint{2112.05510}.

\bibitem{Greene2023}
\bibinfo{author}{{Greene}, T.~P.} \emph{et~al.}
\newblock \bibinfo{title}{{Thermal emission from the Earth-sized exoplanet TRAPPIST-1 b using JWST}}.
\newblock \emph{\bibinfo{journal}{\nat}} \textbf{\bibinfo{volume}{618}}, \bibinfo{pages}{39--42} (\bibinfo{year}{2023}).
\newblock \eprint{2303.14849}.

\bibitem{Lim2023}
\bibinfo{author}{{Lim}, O.} \emph{et~al.}
\newblock \bibinfo{title}{{Atmospheric Reconnaissance of TRAPPIST-1 b with JWST/NIRISS: Evidence for Strong Stellar Contamination in the Transmission Spectra}}.
\newblock \emph{\bibinfo{journal}{\apjl}} \textbf{\bibinfo{volume}{955}}, \bibinfo{pages}{L22} (\bibinfo{year}{2023}).
\newblock \eprint{2309.07047}.

\bibitem{Zieba2023}
\bibinfo{author}{{Zieba}, S.} \emph{et~al.}
\newblock \bibinfo{title}{{No thick carbon dioxide atmosphere on the rocky exoplanet TRAPPIST-1 c}}.
\newblock \emph{\bibinfo{journal}{\nat}} \textbf{\bibinfo{volume}{620}}, \bibinfo{pages}{746--749} (\bibinfo{year}{2023}).
\newblock \eprint{2306.10150}.

\bibitem{Lincowski2023}
\bibinfo{author}{{Lincowski}, A.~P.} \emph{et~al.}
\newblock \bibinfo{title}{{Potential Atmospheric Compositions of TRAPPIST-1 c Constrained by JWST/MIRI Observations at 15 {\ensuremath{\mu}}m}}.
\newblock \emph{\bibinfo{journal}{\apjl}} \textbf{\bibinfo{volume}{955}}, \bibinfo{pages}{L7} (\bibinfo{year}{2023}).
\newblock \eprint{2308.05899}.

\bibitem{Howard2023}
\bibinfo{author}{{Howard}, W.~S.} \emph{et~al.}
\newblock \bibinfo{title}{{Characterizing the Near-infrared Spectra of Flares from TRAPPIST-1 during JWST Transit Spectroscopy Observations}}.
\newblock \emph{\bibinfo{journal}{\apj}} \textbf{\bibinfo{volume}{959}}, \bibinfo{pages}{64} (\bibinfo{year}{2023}).
\newblock \eprint{2310.03792}.

\bibitem{Moran2023}
\bibinfo{author}{{Moran}, S.~E.} \emph{et~al.}
\newblock \bibinfo{title}{{High Tide or Riptide on the Cosmic Shoreline? A Water-rich Atmosphere or Stellar Contamination for the Warm Super-Earth GJ 486b from JWST Observations}}.
\newblock \emph{\bibinfo{journal}{\apjl}} \textbf{\bibinfo{volume}{948}}, \bibinfo{pages}{L11} (\bibinfo{year}{2023}).
\newblock \eprint{2305.00868}.

\bibitem{rackham2018}
\bibinfo{author}{{Rackham}, B.~V.}, \bibinfo{author}{{Apai}, D.} \& \bibinfo{author}{{Giampapa}, M.~S.}
\newblock \bibinfo{title}{{The Transit Light Source Effect: False Spectral Features and Incorrect Densities for M-dwarf Transiting Planets}}.
\newblock \emph{\bibinfo{journal}{\apj}} \textbf{\bibinfo{volume}{853}}, \bibinfo{pages}{122} (\bibinfo{year}{2018}).
\newblock \eprint{1711.05691}.

\bibitem{Rackham2019A}
\bibinfo{author}{{Rackham}, B.~V.}, \bibinfo{author}{{Apai}, D.} \& \bibinfo{author}{{Giampapa}, M.~S.}
\newblock \bibinfo{title}{{The Transit Light Source Effect. II. The Impact of Stellar Heterogeneity on Transmission Spectra of Planets Orbiting Broadly Sun-like Stars}}.
\newblock \emph{\bibinfo{journal}{\aj}} \textbf{\bibinfo{volume}{157}}, \bibinfo{pages}{96} (\bibinfo{year}{2019}).
\newblock \eprint{1812.06184}.

\bibitem{Witzke2021}
\bibinfo{author}{{Witzke}, V.} \emph{et~al.}
\newblock \bibinfo{title}{{MPS-ATLAS: A fast all-in-one code for synthesising stellar spectra}}.
\newblock \emph{\bibinfo{journal}{\aap}} \textbf{\bibinfo{volume}{653}}, \bibinfo{pages}{A65} (\bibinfo{year}{2021}).
\newblock \eprint{2105.13611}.

\bibitem{Rustamkulov2023}
\bibinfo{author}{{Rustamkulov}, Z.} \emph{et~al.}
\newblock \bibinfo{title}{{Early Release Science of the exoplanet WASP-39b with JWST NIRSpec PRISM}}.
\newblock \emph{\bibinfo{journal}{\nat}} \textbf{\bibinfo{volume}{614}}, \bibinfo{pages}{659--663} (\bibinfo{year}{2023}).
\newblock \eprint{2211.10487}.

\bibitem{Zhang2018}
\bibinfo{author}{{Zhang}, Z.}, \bibinfo{author}{{Zhou}, Y.}, \bibinfo{author}{{Rackham}, B.~V.} \& \bibinfo{author}{{Apai}, D.}
\newblock \bibinfo{title}{{The Near-infrared Transmission Spectra of TRAPPIST-1 Planets b, c, d, e, f, and g and Stellar Contamination in Multi-epoch Transit Spectra}}.
\newblock \emph{\bibinfo{journal}{The Astronomical Journal}} \textbf{\bibinfo{volume}{156}}, \bibinfo{pages}{178} (\bibinfo{year}{2018}).
\newblock \eprint{1802.02086}.

\bibitem{Morley2017}
\bibinfo{author}{{Morley}, C.~V.}, \bibinfo{author}{{Kreidberg}, L.}, \bibinfo{author}{{Rustamkulov}, Z.}, \bibinfo{author}{{Robinson}, T.} \& \bibinfo{author}{{Fortney}, J.~J.}
\newblock \bibinfo{title}{{Observing the Atmospheres of Known Temperate Earth-sized Planets with JWST}}.
\newblock \emph{\bibinfo{journal}{\apj}} \textbf{\bibinfo{volume}{850}}, \bibinfo{pages}{121} (\bibinfo{year}{2017}).
\newblock \eprint{1708.04239}.

\bibitem{Krissansen2018}
\bibinfo{author}{{Krissansen-Totton}, J.}, \bibinfo{author}{{Garland}, R.}, \bibinfo{author}{{Irwin}, P.} \& \bibinfo{author}{{Catling}, D.~C.}
\newblock \bibinfo{title}{{Detectability of Biosignatures in Anoxic Atmospheres with the James Webb Space Telescope: A TRAPPIST-1e Case Study}}.
\newblock \emph{\bibinfo{journal}{\aj}} \textbf{\bibinfo{volume}{156}}, \bibinfo{pages}{114} (\bibinfo{year}{2018}).
\newblock \eprint{1808.08377}.

\bibitem{Lustig-Yaeger2019}
\bibinfo{author}{{Lustig-Yaeger}, J.}, \bibinfo{author}{{Meadows}, V.~S.} \& \bibinfo{author}{{Lincowski}, A.~P.}
\newblock \bibinfo{title}{{The Detectability and Characterization of the TRAPPIST-1 Exoplanet Atmospheres with JWST}}.
\newblock \emph{\bibinfo{journal}{The Astronomical Journal}} \textbf{\bibinfo{volume}{158}}, \bibinfo{pages}{27} (\bibinfo{year}{2019}).
\newblock \eprint{1905.07070}.

\bibitem{Fauchez2019}
\bibinfo{author}{{Fauchez}, T.~J.} \emph{et~al.}
\newblock \bibinfo{title}{{Impact of Clouds and Hazes on the Simulated JWST Transmission Spectra of Habitable Zone Planets in the TRAPPIST-1 System}}.
\newblock \emph{\bibinfo{journal}{\apj}} \textbf{\bibinfo{volume}{887}}, \bibinfo{pages}{194} (\bibinfo{year}{2019}).
\newblock \eprint{1911.08596}.

\bibitem{Wunderlich2019}
\bibinfo{author}{{Wunderlich}, F.} \emph{et~al.}
\newblock \bibinfo{title}{{Detectability of atmospheric features of Earth-like planets in the habitable zone around M dwarfs}}.
\newblock \emph{\bibinfo{journal}{\aap}} \textbf{\bibinfo{volume}{624}}, \bibinfo{pages}{A49} (\bibinfo{year}{2019}).
\newblock \eprint{1905.02560}.

\bibitem{Gialluca2021}
\bibinfo{author}{{Gialluca}, M.~T.}, \bibinfo{author}{{Robinson}, T.~D.}, \bibinfo{author}{{Rugheimer}, S.} \& \bibinfo{author}{{Wunderlich}, F.}
\newblock \bibinfo{title}{{Characterizing Atmospheres of Transiting Earth-like Exoplanets Orbiting M Dwarfs with James Webb Space Telescope}}.
\newblock \emph{\bibinfo{journal}{\pasp}} \textbf{\bibinfo{volume}{133}}, \bibinfo{pages}{054401} (\bibinfo{year}{2021}).
\newblock \eprint{2101.04139}.

\bibitem{Faria2022}
\bibinfo{author}{{Faria}, J.~P.} \emph{et~al.}
\newblock \bibinfo{title}{{A candidate short-period sub-Earth orbiting Proxima Centauri}}.
\newblock \emph{\bibinfo{journal}{\aap}} \textbf{\bibinfo{volume}{658}}, \bibinfo{pages}{A115} (\bibinfo{year}{2022}).
\newblock \eprint{2202.05188}.

\bibitem{Rackham2023}
\bibinfo{author}{{Rackham}, B.~V.} \& \bibinfo{author}{{de Wit}, J.}
\newblock \bibinfo{title}{{Towards robust corrections for stellar contamination in JWST exoplanet transmission spectra}}.
\newblock \emph{\bibinfo{journal}{arXiv e-prints}} \bibinfo{pages}{arXiv:2303.15418} (\bibinfo{year}{2023}).
\newblock \eprint{2303.15418}.

\bibitem{Berardo2024}
\bibinfo{author}{{Berardo}, D.}, \bibinfo{author}{{de Wit}, J.} \& \bibinfo{author}{{Rackham}, B.~V.}
\newblock \bibinfo{title}{{Empirically Constraining the Spectra of Stellar Surface Features Using Time-resolved Spectroscopy}}.
\newblock \emph{\bibinfo{journal}{\apjl}} \textbf{\bibinfo{volume}{961}}, \bibinfo{pages}{L18} (\bibinfo{year}{2024}).
\newblock \eprint{2307.04785}.

\bibitem{Vogler2005}
\bibinfo{author}{{V{\"o}gler}, A.} \emph{et~al.}
\newblock \bibinfo{title}{{Simulations of magneto-convection in the solar photosphere. Equations, methods, and results of the MURaM code}}.
\newblock \emph{\bibinfo{journal}{\aap}} \textbf{\bibinfo{volume}{429}}, \bibinfo{pages}{335--351} (\bibinfo{year}{2005}).

\bibitem{Morris2018}
\bibinfo{author}{Morris, B.~M.} \emph{et~al.}
\newblock \bibinfo{title}{Non-detection of contamination by stellar activity in the spitzer transit light curves of {TRAPPIST}-1}.
\newblock \emph{\bibinfo{journal}{The Astrophysical Journal}} \textbf{\bibinfo{volume}{863}}, \bibinfo{pages}{L32} (\bibinfo{year}{2018}).
\newblock \urlprefix\url{https://doi.org/10.3847/2041-8213/aad8aa}.

\bibitem{Krissansen2023}
\bibinfo{author}{{Krissansen-Totton}, J.}
\newblock \bibinfo{title}{{Implications of Atmospheric Nondetections for Trappist-1 Inner Planets on Atmospheric Retention Prospects for Outer Planets}}.
\newblock \emph{\bibinfo{journal}{\apjl}} \textbf{\bibinfo{volume}{951}}, \bibinfo{pages}{L39} (\bibinfo{year}{2023}).
\newblock \eprint{2306.05397}.

\bibitem{Redfield2024}
\bibinfo{author}{{Redfield}, S.} \emph{et~al.}
\newblock \bibinfo{title}{{Report of the Working Group on Strategic Exoplanet Initiatives with HST and JWST}}.
\newblock \emph{\bibinfo{journal}{arXiv e-prints}} \bibinfo{pages}{arXiv:2404.02932} (\bibinfo{year}{2024}).
\newblock \eprint{2404.02932}.

\bibitem{Luger2021a}
\bibinfo{author}{{Luger}, R.}, \bibinfo{author}{{Foreman-Mackey}, D.}, \bibinfo{author}{{Hedges}, C.} \& \bibinfo{author}{{Hogg}, D.~W.}
\newblock \bibinfo{title}{{Mapping Stellar Surfaces. I. Degeneracies in the Rotational Light-curve Problem}}.
\newblock \emph{\bibinfo{journal}{\aj}} \textbf{\bibinfo{volume}{162}}, \bibinfo{pages}{123} (\bibinfo{year}{2021}).
\newblock \eprint{2102.00007}.

\bibitem{Luger2021b}
\bibinfo{author}{{Luger}, R.}, \bibinfo{author}{{Foreman-Mackey}, D.} \& \bibinfo{author}{{Hedges}, C.}
\newblock \bibinfo{title}{{Mapping Stellar Surfaces. II. An Interpretable Gaussian Process Model for Light Curves}}.
\newblock \emph{\bibinfo{journal}{\aj}} \textbf{\bibinfo{volume}{162}}, \bibinfo{pages}{124} (\bibinfo{year}{2021}).
\newblock \eprint{2102.01697}.

\bibitem{Luger2022}
\bibinfo{author}{{Luger}, R.} \emph{et~al.}
\newblock \bibinfo{title}{{Mapping stellar surfaces III: An Efficient, Scalable, and Open-Source Doppler Imaging Model}}.
\newblock \emph{\bibinfo{journal}{arXiv e-prints}} \bibinfo{pages}{arXiv:2110.06271} (\bibinfo{year}{2021}).
\newblock \eprint{2110.06271}.

\bibitem{Mallonn2018}
\bibinfo{author}{{Mallonn}, M.} \emph{et~al.}
\newblock \bibinfo{title}{{GJ 1214: Rotation period, starspots, and uncertainty on the optical slope of the transmission spectrum}}.
\newblock \emph{\bibinfo{journal}{\aap}} \textbf{\bibinfo{volume}{614}}, \bibinfo{pages}{A35} (\bibinfo{year}{2018}).
\newblock \eprint{1803.05677}.

\bibitem{Rosich2020}
\bibinfo{author}{{Rosich}, A.} \emph{et~al.}
\newblock \bibinfo{title}{{Correcting for chromatic stellar activity effects in transits with multiband photometric monitoring: application to WASP-52}}.
\newblock \emph{\bibinfo{journal}{\aap}} \textbf{\bibinfo{volume}{641}}, \bibinfo{pages}{A82} (\bibinfo{year}{2020}).
\newblock \eprint{2007.00573}.

\bibitem{Perger2023}
\bibinfo{author}{{Perger}, M.} \emph{et~al.}
\newblock \bibinfo{title}{{A machine learning approach for correcting radial velocities using physical observables}}.
\newblock \emph{\bibinfo{journal}{\aap}} \textbf{\bibinfo{volume}{672}}, \bibinfo{pages}{A118} (\bibinfo{year}{2023}).
\newblock \eprint{2301.12872}.

\bibitem{Reyle2021}
\bibinfo{author}{{Reyl{\'e}}, C.} \emph{et~al.}
\newblock \bibinfo{title}{{The 10 parsec sample in the Gaia era}}.
\newblock \emph{\bibinfo{journal}{\aap}} \textbf{\bibinfo{volume}{650}}, \bibinfo{pages}{A201} (\bibinfo{year}{2021}).
\newblock \eprint{2104.14972}.

\bibitem{Hansen1994}
\bibinfo{author}{{Hansen}, C.~J.} \& \bibinfo{author}{{Kawaler}, S.~D.}
\newblock \emph{\bibinfo{title}{{Stellar Interiors. Physical Principles, Structure, and Evolution}}}, \bibinfo{Publisher}{Springer}, \bibinfo{City}{New York, NY} (\bibinfo{year}{1994}). 
\newblock \urlprefix\url{https://ui.adsabs.harvard.edu/abs/1994sipp.book.....H}.

\bibitem{Kroupa2001}
\bibinfo{author}{{Kroupa}, P.}
\newblock \bibinfo{title}{{On the variation of the initial mass function}}.
\newblock \emph{\bibinfo{journal}{\mnras}} \textbf{\bibinfo{volume}{322}}, \bibinfo{pages}{231--246} (\bibinfo{year}{2001}).
\newblock \eprint{astro-ph/0009005}.

\bibitem{Kempton2018}
\bibinfo{author}{{Kempton}, E. M.~R.} \emph{et~al.}
\newblock \bibinfo{title}{{A Framework for Prioritizing the TESS Planetary Candidates Most Amenable to Atmospheric Characterization}}.
\newblock \emph{\bibinfo{journal}{\pasp}} \textbf{\bibinfo{volume}{130}}, \bibinfo{pages}{114401} (\bibinfo{year}{2018}).
\newblock \eprint{1805.03671}.

\bibitem{speclib}
\bibinfo{author}{{Rackham}, B.~V.}
\newblock \bibinfo{title}{{speclib}} (\bibinfo{year}{2023})
\newblock \urlprefix\url{https://ui.adsabs.harvard.edu/abs/2023zndo...7868050R}.

\bibitem{Morris2020}
\bibinfo{author}{{Morris}, B.}
\newblock \bibinfo{title}{{fleck: Fast approximate light curves for starspot rotational modulation}}.
\newblock \emph{\bibinfo{journal}{The Journal of Open Source Software}} \textbf{\bibinfo{volume}{5}}, \bibinfo{pages}{2103} (\bibinfo{year}{2020}).

\bibitem{Niraula2022}
\bibinfo{author}{{Niraula}, P.} \emph{et~al.}
\newblock \bibinfo{title}{{The impending opacity challenge in exoplanet atmospheric characterization}}.
\newblock \emph{\bibinfo{journal}{Nature Astronomy}} \textbf{\bibinfo{volume}{6}}, \bibinfo{pages}{1287-1295} (\bibinfo{year}{2022}).

\end{thebibliography}

\end{document}